\documentclass{article}

\usepackage{amsthm,amsmath,amssymb,arydshln,url}
\RequirePackage[colorlinks,citecolor=blue,urlcolor=blue]{hyperref}
\usepackage[english]{babel}
\usepackage{graphicx}
\usepackage{multirow,multicol,color,rotating,enumitem,colortbl}
\usepackage{float} 
\usepackage{natbib}
\newcommand{\PI}{\boldsymbol{\theta}}
\newcommand{\Nbtab}{M}

\newcommand{\Nbind}{n}
\newcommand{\nbind}{i}
\newcommand{\Nbvar}{p}
\newcommand{\Nbfix}{q}
\newcommand{\Nbrand}{q'}
\newcommand{\nbvar}{j}
\newcommand{\nbcol}{j}
\newcommand{\Nbout}{r}
\newcommand{\Nbsim}{T}

\newcommand{\Nbclust}{{K}}
\newcommand{\nbclust}{{k}}

\newcommand{\bfW}{{\textbf{W}}}

\newcommand{\bfY}{{\textbf{Y}}}
\newcommand{\bfy}{{\textbf{y}}}
\newcommand{\bfZ}{{\textbf{Z}}}
\newcommand{\bfz}{{\textbf{z}}}
\newcommand{\bfbeta}{\boldsymbol{\beta}}
\newcommand{\bfPsi}{\boldsymbol{\Psi}}
\newcommand{\bfSigma}{\boldsymbol{\Sigma}}

\newcommand{\bfLambda}{\boldsymbol{\Lambda}}
\newcommand{\bfun}{\boldsymbol{1}}
\definecolor{mypink}{cmyk}{0, 0.7808, 0.4429, 0.1412}

\newlength{\sidewaysclen}
\newcommand{\sidewaysc}[1]{\raisebox{\dimexpr-0.5\sidewaysclen}{%
  \begin{sideways}\begin{minipage}{\sidewaysclen}%
    \centering #1\end{minipage}\end{sideways}}}

  \makeatletter
\def\@xfootnote[#1]{%
  \protected@xdef\@thefnmark{#1}%
  \@footnotemark\@footnotetext}
\makeatother

\begin{document}
\title{Multiple imputation for multilevel data with continuous and binary variables}

\maketitle

Vincent Audigier$^{1,2,3}$ \footnote[*]{\texttt{vincent.audigier@univ-paris-diderot.fr}}, Ian R. White$^{4,5}$,
Shahab Jolani$^{6}$,
Thomas P. A. Debray$^{7,8}$,
Matteo Quartagno$^{9}$,
James Carpenter$^{9}$,
Stef van Buuren$^{10}$,
Matthieu Resche-Rigon$^{1,2,3}$\vspace{1cm}
\begin{small}
\begin{enumerate}
\item Service de Biostatistique et Information M\'edicale, H\^opital Saint-Louis, AP-HP, Paris, France
\item Universit\'e Paris Diderot - Paris 7, Sorbonne Paris Cit\'e, UMR-S 1153, Paris, France 
\item INSERM, UMR 1153, Equipe ECSTRA, H\^opital Saint-Louis, Paris
\item MRC Biostatistics Unit, Cambridge Institute of Public Health, U.K.
\item MRC Clinical Trials Unit at UCL, London, U.K.
\item Department of Methodology and Statistics, School of Public Health and Primary Care, Maastricht University, Maastricht, The Netherlands
\item Julius Center for Health Sciences and Primary Care, University Medical Center Utrecht, Utrecht, The Netherlands
\item Cochrane Netherlands, University Medical Center Utrecht, Utrecht,  The Netherlands
\item Department of Medical Statistics, London School of Hygiene \& Tropical Medicine, London, U.K.
\item Department of Statistics, TNO Prevention and Health, Leiden, The Netherlands
\end{enumerate}
\end{small}

\newpage
\begin{abstract}
We present and compare multiple imputation methods for multilevel continuous and binary data where variables are systematically and sporadically missing.

The methods are compared from a theoretical point of view and through an extensive simulation study motivated by a real dataset comprising multiple studies. Simulations are reproducible. The comparisons show why these multiple imputation methods are the most appropriate to handle missing values in a multilevel setting and why their relative performances can vary according to the missing data pattern, the multilevel structure and the type of missing variables.

This study shows that valid inferences can only be obtained if the dataset gathers a large number of clusters. In addition, it highlights that heteroscedastic MI methods provide more accurate inferences than homoscedastic methods, which should be reserved for data with few individuals per cluster. Finally, the method of \cite{Quartagno15} appears generally accurate for binary variables, the method of \cite{Resche16} with large clusters, and the approach of \cite{Jolani15} with small clusters.
\end{abstract}


\section{Introduction}
When individual observations are nested in clusters, statistical analyses generally need to reflect this structure; this is usually referred to as a multilevel structure, where individuals constitute the lower level, and clusters the higher level. This situation often arises in fields like survey research, educational science, sociology, geography, psychology and clinical studies among others.

Missing data affect nearly any dataset in most of these fields, and multilevel data present specific patterns of missing values. Some variables may be fully non-observed for some clusters because they were not measured, or because they were not defined consistently across clusters. \cite{Resche13} named such missing data patterns \textit{systematically missing values}. Nowadays, systematically missing values are becoming increasingly common because of a greater availability of data coming from several sources, including different sets of variables \citep{Riley16}. Since systematically missing values are related to the multilevel structure, they affect any field where multilevel data occur. Examples of multilevel data with systematically missing values include \cite{Bos03}, \cite{Mullis03} or \cite{Blossfeld11} in educational sciences, \cite{great}, \cite{Kunkel17} in medicine, \cite{Carrig15} in sociology. As opposed to systematically missing values, \textit{sporadically missing values} are missing data specific to each individual observation, like in the common single-level framework. It is often the case that both types of missing data occur in a multilevel dataset. For instance, in educational research, questionnaires may be too long to be administered to all students and, therefore, only a subset of items may be asked to each class, leading to systematically missing values. In addition, some students in each class may not answer some questions, leading to sporadically missing values. 

Multiple imputation (MI) is a common strategy to deal with missing values in statistical analysis \citep{Schafer97,Rubin87,Little02}. It involves firstly specifying a distribution in accordance with the data, the \textit{imputation model}, under which $\Nbtab$ imputations are drawn from their posterior predictive distribution given the observed values. Thus, $\Nbtab$ complete datasets are generated. Secondly, a standard statistical analysis is performed on each imputed dataset, leading to $\Nbtab$ estimates of the \textit{analysis model}'s parameters. Finally, the estimates are pooled according to Rubin's rules \citep{Rubin87}. The standard assumption when using MI is ignorability \citep[p.~11]{Schafer97}, implying that missing values occur \textit{at random} \citep{Rubin76}, \textit{i.e.} the probability of missingness depends solely on observed data. 
Several MI methods have been proposed, differing mainly in the assumed form of the imputation model; among methods assuming a parametric imputation model, two strategies are the most common. We refer to joint modelling (JM) imputation when a multivariate joint distribution is specified for all variables. Alternatively, we refer to fully conditional specification (FCS) when a conditional distribution is defined for each incomplete variable \citep{vanBuuren06, Raghunathan01}.

In the standard statistical framework where data are complete, multilevel data induce non-independence between observations and require dedicated analysis models accounting for this source of dependency. The linear mixed effect model is one of such models. In the same way, with missing values, imputation models need to take into account dependency between observations, or otherwise the variance of prediction of missing values cannot be properly reflected. Indeed, when an incomplete variable is part of a linear mixed effect analysis model, but it is imputed ignoring the multilevel structure, the imputed values are too close to their expectation. 
Thus, biases can occur even when applying appropriate statistical methods on inappropriately imputed data.

To account for the multilevel structure with sporadically missing values, imputation models are generally based on regression models including a fixed or a random intercept for cluster \citep{Drechsler15}. Methods using a fixed intercept treat the identifier of each cluster as a dummy variable. They are generally parametric, using normal regression for instance, but imputation according to semi-parametric method can also be relevant for complex datasets \citep{Vink15,Little88}. However, using a random intercept is generally preferable because fixed intercept inflates the true variability between clusters \citep{Andridge11, Graham12}. MI methods using a random intercept include JM-pan \citep{Yucel02}, JM-jomo \citep{Quartagno15}, JM-REALCOM \citep{Goldstein07,Goldstein09,Carpenter13}, JM-Mplus \citep{Asparouhov10}, JM-RCME \citep{Yucel11}, FCS-pan \citep{Yucel02}, FCS-blimp \citep{Enders17}, FCS-2lnorm \citep{vanBuuren10}, FCS-GLM  \citep{Jolani15, Jolani17}, FCS-2stage  \citep{Resche16}. These methods differ by the form of the imputation model (joint or not), but also by their ability to account for different types of variables (continuous, binary, or others). In particular, JM-pan, JM-RCME, FCS-pan, FCS-2lnorm are not dedicated to binary variables.

Systematically missing values imply identifiability issues for imputation models which include a fixed intercept for cluster. Thus, only methods using random effects are appealing. Most of the first MI methods proposed to impute multilevel data were based on random intercept models, but systematically missing values were not considered. Therefore, not all of these methods are tailored for the imputation of such missing data. Table \ref{tablemethod} summarizes the MI methods available for multilevel data.


\begin{table}[H]
\begin{center}
\caption{Summary of MI methods' properties for multilevel data based on random intercept (JM-pan \citep{Yucel02}, JM-REALCOM \citep{Goldstein07,Goldstein09,Carpenter13}, JM-jomo \citep{Quartagno15}, JM-Mplus \citep{Asparouhov10}, JM-RCME \citep{Yucel11}, FCS-pan \citep{Yucel02},  FCS-blimp \citep{Enders17} FCS-2lnorm \citep{vanBuuren10}, FCS-GLM \citep{Jolani15}, FCS-2stage \citep{Resche16})\label{tablemethod}}
\begin{tabular}{p{2.3cm}p{1.3cm}p{1.5cm}p{1.5cm}p{1.3cm}p{2.8cm}}
 \hline Method ~~~~~~~~ (form - name)						&	\multicolumn{4}{c}{Handles missing data:}						&	Coded in R \citep{Rsoft}\\ \cline{2-5}
										&{Sporadic?}&{Systematic?}& {in continuous variable?}&{in binary variable?}											&\\ \hline
JM-pan			&yes&yes&yes&no&yes, package \textit{pan}\\
JM-REALCOM			&yes&yes&yes&yes&no \\
JM-jomo				&yes&yes&yes&yes&yes, package \textit{jomo}	\\
JM-Mplus			&yes&yes&yes&yes&no\\
JM-RCME				&yes&yes&yes&no&no\\
FCS-pan				&yes&yes&yes&no&yes, package \textit{mice}\\
FCS-blimp			&yes&yes&yes&yes&no\\
FCS-2lnorm			&yes&no&yes&no&yes, package \textit{mice}\\
FCS-GLM				&yes\protect\footnotemark \addtocounter{footnote}{-1}&yes&yes&yes&yes, add-on for \textit{mice}\\
FCS-2stage (REML or MM)&yes&yes&yes&yes\protect\footnotemark&yes, add-on for \textit{mice}\\ \hline
\end{tabular}
\begin{flushright}$^1$ using variant reported in this paper\end{flushright}
\end{center}
\end{table}

In this paper, we compare the most relevant MI methods for dealing with clustered datasets with systematically and sporadically missing variables, continuous and binary. Among JM methods for multilevel data, we focus on the JM-jomo method proposed in \cite{Quartagno15}, which can be seen as a generalisation of JM-REALCOM and JM-Mplus, while among FCS methods, we focus on the FCS-GLM method presented in \cite{Jolani15} and on the FCS-2stage method proposed in \cite{Resche16}. We do not focus on FCS-blimp which can be seen as a univariate version of JM-jomo. 

The paper is organised as follows. Firstly, we present the three MI methods for handling multilevel data with systematically and sporadically missing values (Section \ref{MImethods}). Both FCS methods have theoretical deficiencies in this general setting, so we propose improvements for them: accounting for binary variables in FCS-2stage, and accounting for continuous sporadically missing values in FCS-GLM. Secondly, these MI methods are compared through a simulation study (Section \ref{Simulation}). Thirdly, MI methods are applied to a real data analysis (Section \ref{Application}). Finally, practical recommendations are provided (Section \ref{Conclusion}).

\section{Multiple imputation for multilevel continuous and binary data}
\label{MImethods}
\subsection{Methods}

\subsubsection{Univariate missing data pattern}
\label{univdescript}
Random variables will be indicated in italics, while fixed values will be denoted in roman letters. Vectors will be in lower case, while matrices will be in upper case. Let $\bfY_{\Nbind \times \Nbvar}=\left(\bfy_1,\ldots,\bfy_{\Nbvar}\right)$ be an incomplete data matrix for $\Nbind$ individuals in rows and $\Nbvar$ variables in columns. Let $\nbind$ be the index for the individuals ($1\leq\nbind\leq\Nbind$) and $\nbvar$ for the columns ($1\leq \nbvar\leq\Nbvar$). 
$\bfY$ is stratified into $\Nbclust$ clusters of size $\Nbind_\nbclust$ where $\nbclust$ denotes the index for the cluster $\left(1\leq \nbclust\leq\Nbclust\right)$. $\bfy_{\nbvar\nbclust}$ denotes the $\Nbind_{\nbclust}$-vector corresponding to the vector $\bfy_{\nbvar}$ restricted to individuals within cluster $\nbclust$.
Let $\left(\bfy_{\nbvar}^{\text{obs}},\bfy_{\nbvar}^{\text{miss}}\right)$ be the missing and observed parts of $\bfy_{\nbvar}$ and let $\bfY^{\text{obs}}=\left(\bfy_1^{\text{obs}},\ldots,\bfy_{\Nbvar}^{\text{obs}}\right)$ and $\bfY^{\text{miss}}=\left(\bfy_1^{\text{miss}},\ldots,\bfy_{\Nbvar}^{\text{miss}}\right)$.

In order to propose a unified presentation of the three MI methods, we assume in this section that the variable $y=y_\Nbvar$ is the only incomplete variable and is continuous. Extension to several incomplete variables, continuous or binary, will be discussed in the next section.\\

The imputation step in MI aims to draw missing values from their predictive distribution $P\left(Y^{\text{miss}}\vert Y^{\text{obs}}\right)$. To achieve this goal, an imputation model with parameter $\PI$ is specified and realisations of the predictive distribution of missing values can be obtained by: \begin{enumerate}[label=Step (\arabic*)]
\item drawing $\PI$ from $P\left(\PI \vert Y^{\text{obs}}  \right)$, its posterior distribution \label{MIstep1}
\item drawing missing data according to $P\left(Y^{\text{miss}}\vert Y^{\text{obs}}, \PI \right)$, their predictive distribution for a given $\PI$. \label{MIstep2}
\end{enumerate}

For a single continuous incomplete variable, the posterior distribution can be specified by letting $\PI=\left(\bfbeta, \bfPsi, \left(\bfSigma_{\nbclust}\right)_{1\leq\nbclust\leq\Nbclust}\right)$ be the parameters of a linear mixed effects model:
\begin{eqnarray}\label{glm}
y_{\nbclust}&=&\bfZ_{\nbclust}\bfbeta+\bfW_{\nbclust}b_{\nbclust}+\varepsilon_{\nbclust}\\
b_{\nbclust}&\sim& \mathcal{N}\left({\bf{0}},\bfPsi\right) \nonumber\\
\varepsilon_{\nbclust} &\sim& \mathcal{N}\left({\bf{0}},\bfSigma_{\nbclust}\right)\nonumber
\end{eqnarray}
where $y_\nbclust$ denotes the incomplete variable restricted to the cluster $\nbclust$, $\bfZ_{\nbclust}$ $\left(\Nbind_{\nbclust}\times \Nbfix \right)$ and $\bfW_{\nbclust}$  $\left(\Nbind_{\nbclust}\times \Nbrand \right)$ are the known covariate matrices corresponding to two subsets of $\left(\bfy_{1\nbclust},\ldots,\bfy_{\left(\Nbvar-1\right)\nbclust}\right)$, $\bfbeta$ is the $\Nbfix$ vector of regression coefficients of fixed effects, $b_{\nbclust}$ is the $\Nbrand$ vector of random effects for cluster $\nbclust$, $\bfPsi$  $\left(\Nbrand\times \Nbrand\right)$ is the between cluster variance matrix, $\bfSigma_\nbclust=\sigma^2_{\nbclust}{{\mathbb{I}}}_{\Nbind_{\nbclust}}$ $\left(\Nbind_{\nbclust}\times \Nbind_{\nbclust}\right)$ is the variance matrix within cluster $\nbclust$. Model (\ref{glm}) is the imputation model used in FCS-GLM and FCS-2stage and potentially in JM-jomo for the case of a univariate missing data pattern.\\

Drawing parameters of the imputation model from their posterior distribution (\ref{MIstep1}) can be achieved by several approaches \citep[p200-222]{Little02}. A first approach uses explicit Bayesian modelling of (\ref{glm}), specifying a prior distribution for $\PI$ and drawing from its posterior distribution. This approach is used in FCS-GLM and in JM-jomo: FCS-GLM uses a non-informative Jeffreys prior distribution, while JM-jomo uses a conjugate prior distribution.

A second approach uses the asymptotic distribution of a frequentist estimator of $\PI$\label{asympmethod}. More precisely, the parameters of this distribution are estimated from the data, and a value of $\PI$ is drawn from this asymptotic distribution. FCS-2stage is based on this principle: the estimator used is called the \textit{two-stage estimator} in IPD meta-analysis \citep{Simmonds05,Riley08}. It is also possible to use the Maximum Likelihood (ML) estimator \citep{Resche13}, but the two-stage estimator has the advantage of being easier and quicker to compute than the ML estimator for linear mixed effects models.

When the variable $y$ is only sporadically missing, the posterior distribution of the parameters only involves individuals that are observed \citep[p. 165]{Rubin87}, so that both approaches easily handle missing data. However, systematically missing values complicate \ref{MIstep1} for both approaches. Using Bayesian modelling, simulating the posterior distribution of $\PI$ generally requires a Gibbs sampler \citep{Geman84}, but the posterior distribution of $\bfSigma_{\nbclust}$ cannot be updated from the data at each iteration for systematically missing clusters. Similarly, $\bfSigma_{\nbclust}$ cannot be estimated from observed data by using the asymptotic method. Thus, MI methods developed for sporadically missing data cannot be directly used to impute systematically missing data. 
The problem is overcome by assuming a distribution across $\left(\bfSigma_{\nbclust}\right)_{1\leq \nbclust\leq\Nbclust}$, as proposed in JM-jomo and in FCS-2stage, or by assuming $\bfSigma_{\nbclust}=\bfSigma$ for all $\nbclust$, as proposed in FCS-GLM.\\

To draw missing values according to the parameters drawn at \ref{MIstep1}, missing values are predicted according to model (\ref{glm}) and Gaussian noise is added to the prediction (\ref{MIstep2}). However, the random coefficients are not strictly parameters of this model and, therefore, are not directly given by \ref{MIstep1}. Thus, to obtain realisations for $\left(b_\nbclust\right)_{1\leq\nbclust\leq\Nbclust}$, each random coefficient is drawn from its distribution conditional on $y^{\text{obs}}_{\nbclust}$ and the parameters generated from \ref{MIstep1}. Imputation can then be performed. If data are sporadically missing, these conditional distributions are derived by classic calculation for Gaussian vectors; if data are systematically missing, random coefficients are drawn from their marginal distribution.\\

From this unified presentation, we now present the methods for several incomplete variables, which can also be binary.

\subsubsection{Multivariate missing data pattern}
 \label{MImethod}

\paragraph{JM-jomo \citep{Quartagno15}\label{jomomethod}.}
To multiply impute multilevel data with several incomplete continuous variables, JM approaches are based on the multivariate version of model (\ref{glm}) where covariate matrices are matrices ($\nbind_\nbclust \times \Nbvar$) of ones:
\begin{eqnarray}
\label{glmmulti}
Y_\nbclust&=&\bfun\bfbeta+\bfun b_{\nbclust}+\varepsilon_{\nbclust}\\
b_{\nbclust}^{V}&\sim& \mathcal{N}\left(0,\bfPsi\right) \nonumber\\
\varepsilon_{\nbclust}^{V} &\sim& \mathcal{N}\left(0,\bfSigma_{\nbclust}\right)\nonumber
\end{eqnarray}
$\bfbeta$ is the $1 \times \Nbout$ matrix of regression coefficients of fixed effects, and
$b_{\nbclust}$ is the $1\times \Nbout$ matrix of random effects.
The superscript $V$ indicates the vectorisation of a matrix by stacking its columns.
$\bfPsi$  $\left(\Nbout\times \Nbout\right)$ is the between cluster variance matrix and $\bfSigma_\nbclust$ $\left(\Nbout\Nbind_{\nbclust}\times \Nbout\Nbind_{\nbclust}\right)$ is the block diagonal variance matrix within cluster $\nbclust$.

Note that model (\ref{glmmulti}) includes all variables on the left hand side of imputation model ($Y_\nbclust=\left(Y_\nbclust^{\text{miss}},Y_\nbclust^{\text{obs}}\right)$). Other modelling could be considered by including complete variables can be included on the left or right hand side. The proposed model has the advantage to limit overfitting when the number of variables is small compared to the number of individuals \cite{Quartagno15}.

To perform MI according to this imputation model, a Bayesian approach is used with the following independent prior distributions for $\PI=\left(\bfbeta,\bfPsi,\left(\bfSigma_{\nbclust}\right)_{1\leq\nbclust\leq\Nbclust}\right)$:
\begin{eqnarray}
\bfbeta\propto 1\\
\bfPsi^{-1}\sim \mathcal{W}\left(\nu_1,\bfLambda_1\right)\\
\bfSigma_{\nbclust}^{-1}\vert\nu_2,\bfLambda_2 \sim \mathcal{W}\left(\nu_2,\bfLambda_2\right) \text{for all } {1\leq\nbclust\leq\Nbclust} \\
\nu_2 \sim \chi^2\left(\eta\right),\ \bfLambda_2^{-1} \sim \mathcal{W}\left(\nu_3,\bfLambda_3\right)\label{eqnjm}
\end{eqnarray}
where $\mathcal{W}\left(\nu,\bfLambda\right)$ denotes the Wishart distribution with $\nu$ degrees of freedom and scale matrix  $\bfLambda$, and $\eta$ denotes the degrees of freedom of the chi-squared distribution. The prior distributions for the covariance matrices are informative \citep{Gelman06}; to make them as vague as possible, hyperparameters are set as $\nu_1=\Nbout$, $\bfLambda_1=\mathbb{I}_{\Nbout}$, $\nu_3=\Nbout\Nbclust$, $\bfLambda_3=\mathbb{I}_{\Nbout\Nbclust}$ and $\eta=\Nbout\Nbclust$. From this modelling, the posterior distribution of $\PI$ can be derived. We report in Appendix \ref{pd_jomo} the derived posterior distributions as well as the technical details to obtain realisations from them.

In summary, the parameters of the imputation model are drawn from their posterior distribution with a multivariate missing data pattern by using a Data-Augmentation (DA) algorithm \citep{Tanner87}: given current values $\PI^{(\ell)}$ for $\PI$ and ${Y^{\text{miss}}}^{(\ell)}$ for $Y^{\text{miss}}$, the components of $\PI$ are successively updated according to their posterior distribution (Equations \ref{jm1}-\ref{jm3} in Appendix \ref{pd_jomo}) given $\left(Y^{\text{obs}},{Y^{\text{miss}}}^{(\ell)}\right)$, providing $\PI^{(\ell+1)}$. Then, $\PI^{(\ell+1)}$ can be used to generate ${Y^{\text{miss}}}^{(\ell+1)}$ according to model (\ref{glmmulti}). To obtain $\Nbtab$ independent realisations from the posterior distribution, the algorithm is run through a burn-in period (to reach the convergence to the posterior distribution) and then realisations are drawn by spacing them with several iterations (to ensure independence). Note that the number of iterations for the burn-in period and the number of iterations between to realisations need to be carefully checked \citep[p.160-169]{Schafer97}. Moreover, since generating $\PI$ in its predictive distribution using the DA algorithm also requires imputation of missing data, \ref{MIstep1} and \ref{MIstep2} of MI (see Section \ref{univdescript}) are not distinguished here. \\

This method allows imputation of datasets with systematically and sporadically missing values.
In particular, despite systematically missing values, the posterior distribution for $\bfSigma_{\nbclust}$ can be updated at each step of the DA algorithm by considering observed values from other clusters.\\

To deal with binary variables, a probit link and a latent variables framework have been proposed \citep{Goldstein09}. Let $L$ be the set of continuous variables joined with a set of latent variables corresponding to the binary variables, so that $L=\left(L^{\text{miss}}, L^{\text{obs}}
\right)$. 
At the end of each cycle of the DA algorithm, given current parameters $\PI^{(\ell)}$ and random coefficients $b^{(\ell)}$, $L^{\text{miss}}$ is drawn conditionally on $L^{\text{obs}}$.

Like $P\left(Y^{\text{miss}},Y^{\text{obs}}\right)$ for continuous incomplete variables, $P\left(L^{\text{miss}},L^{\text{obs}}\right)$ is a multivariate Gaussian distribution. Thus, drawing missing latent variables consists of drawing $L$ from a Gaussian distribution under the positivity or negativity constraint imposed by observed binary values, which is straightforward \citep[p.96-98]{Carpenter13}. Next binary data from $Y^{\text{miss}}$ are derived from the previously drawn latent variables: the outcome 1 is drawn if the latent variable takes a positive value, and 0 otherwise.\\

The JM-jomo method is an extension of the JM-RCME \citep{Yucel11}, JM-Mplus \citep{Asparouhov10}, JM-REALCOM \citep{Goldstein07,Goldstein09,Carpenter13} and the JM-pan method \citep{Yucel02}; JM-jomo additionally allows for heteroscedasticity of the imputation model and imputation of binary (and more generally categorical) variables, while JM-RCME only handles heteroscedasticity, and JM-REALCOM or JM-Mplus only handle categorical variables. Note that the Bayesian formulation of the imputation model is also very close to those used in FCS-blimp, FCS-pan and FCS-2lnorm \vspace{.5cm}

\paragraph{FCS-GLM \citep{Jolani15}.}
Instead of using a JM approach, fully conditional specification can be used to multiply impute a dataset with several incomplete variables. The principle is to simulate the predictive distribution of the missing values by successively simulating the predictive distribution of the missing values of each incomplete variable conditionally on the other variables. Thus, instead of specifying a joint imputation model as (\ref{glmmulti}), only the conditional distribution of each incomplete variable is required. Compared to JM approaches, FCS approaches make it easier to model complex dependence structures.

\cite{Jolani15} use a FCS approach to perform multiple imputation of systematically missing variables only. For continuous incomplete variables, the conditional imputation model is model (\ref{glm}) assuming homoscedastic error terms, \textit{i.e.} $\sigma_\nbclust=\sigma$ for all $\nbclust$. To draw missing values of $y$ from their predictive distribution, a Bayesian formulation of the univariate linear mixed effects model based on non-informative independent priors is used. Details on the posterior distributions are provided in Appendix \ref{pd_glm}, we only underline that they depend of the maximum likelihood estimates of the imputation model's parameters. 

Imputation of a systematically missing variable $y$ is performed as follows:
\begin{enumerate}[label=Step (\arabic*)]
\item $\PI$ is drawn according to the posterior (equations (\ref{fcs3}-\ref{fcs2}) in Appendix \ref{pd_glm}) \label{fcs4a}
\item $P\left(y^{\text{miss}}\vert Y^{\text{obs}}, \PI \right)$ is simulated by:
\begin{itemize}
\item drawing ${b}_{\nbclust}$ from $\mathcal{N}\left(0,\bfPsi\right)$ for all clusters in $1\leq\nbclust\leq\Nbclust$ where $y_{\nbclust}$ is systematically missing \label{fcs4}
\item drawing $y_{\nbclust}^{\text{miss}}$ from $\mathcal{N}\left(\bfZ_{\nbclust}\bfbeta+\bfW_{\nbclust}b_{\nbclust},\sigma^2\mathbb{I}_{\Nbind_\nbclust}\right)$ for all clusters in $1\leq\nbclust\leq\Nbclust$ where $y_{\nbclust}$ is sporadically missing. \label{fcs5}
\end{itemize}
\end{enumerate}

Binary variables are imputed in the same way by considering a GLMM model with a logit link.\\

FCS-GLM was originally developed to impute systematically missing variables only. We extend it to also impute sporadically missing continuous variable following the rationale of \cite{Resche13}. To achieve this goal, \ref{fcs4a} is essentially the same, the main difference lies in \ref{fcs4}: each $b_{\nbclust}$ is drawn conditionally on $y_{\nbclust}^{\text{obs}}$, instead of being drawn from its marginal distribution. However, when $y$ is a binary variable, this conditional distribution is analytically intractable because of the logit link. Therefore, binary sporadically missing variables are handled as binary systematically missing ones, which can potentially introduce bias and a lack of variability in the imputed values.\\

FCS-2lnorm \citep{vanBuuren10}, FCS-blimp \citep{Enders17} and FCS-pan \citep{Yucel02} also proposed FCS approaches which use a conjugate prior to reflect the posterior distribution of the parameter of the imputation model. FCS-2lnorm is based on the model (\ref{glm}) as conditional imputation model and thus allows heteroscedasticity of errors for continuous variables. However, it cannot be directly applied to systematically missing clusters because of non-identifiability of $(\sigma^2_\nbclust)$ for $1\leq\nbclust \leq \Nbclust$. On the contrary, both others assume homoscedasticity only. \vspace{.5cm}

\paragraph{FCS-2stage \citep{Resche16}.}
FCS-2stage is another FCS method drawing the parameters of the imputation model by using an asymptotic strategy: an estimator is evaluated from the observed data and the posterior distribution is then approximated (\textit{cf} Section \ref{asympmethod}). This estimator is a two-stage estimator \citep{Simmonds05,Riley08}. Often used in IPD meta-analysis, it has the advantage of being quicker to compute than the usual one-stage estimator required for the previous method (through the expressions of posterior distributions).

More precisely, for a continuous incomplete variable $y$, the conditional imputation model (\ref{glm}) is re-written as follows:

\begin{eqnarray}\label{glm2step}
y_{\nbclust}&=&\bfZ_{\nbclust}\left(\bfbeta+b_{\nbclust}\right)+\varepsilon_{\nbclust}\label{model2step}\\
b_{\nbclust}&\sim& \mathcal{N}\left(0,\bfPsi\right) \nonumber\\
\varepsilon_{\nbclust} &\sim& \mathcal{N}\left(0,\sigma^2_{\nbclust}\mathbb{I}_{\Nbind_\nbclust}\right)\nonumber
\end{eqnarray}
Note that for clarity, the method is presented for the case $\bfZ_{\nbclust}=\bfW_{\nbclust}$. Extension to the more general imputation model is given in \cite{Resche16}. The parameter of this model is $\PI=\left(\bfbeta,\bfPsi,\left(\sigma_{\nbclust}\right)_{1\leq\nbclust\leq\Nbclust}\right)$.

To fit the two-stage estimator, at stage one, the ML estimator of a linear model is computed on each available cluster:
\begin{eqnarray}\label{2stage-1}
\widehat{\bfbeta}_{\nbclust}=\left(\bfZ_{\nbclust}^{\top}\bfZ_{\nbclust}\right)^{-1}\bfZ_{\nbclust}^{\top}\bfy_{\nbclust}
\end{eqnarray}
Then, at stage two, the following random effects model is used:
\begin{eqnarray}\label{2stage}
\widehat{\bfbeta}_\nbclust=\bfbeta+b_{\nbclust}+\varepsilon'_{\nbclust}
\end{eqnarray}
with $b_{\nbclust}\sim \mathcal{N}\left(0,\bfPsi\right)$ and $\varepsilon'_{\nbclust}\sim\mathcal{N}\left(0,\sigma_\nbclust^2\left(\bfZ_{\nbclust}\bfZ_{\nbclust}^{\top}\right)^{-1}\right)$.
$\bfbeta$ and $\bfPsi$ may be estimated by REML; alternatively, \cite{Resche16} suggest using the method of moments (MM), which is even faster, especially with high dimensional $\bfbeta$ \citep{derSimonian86,Jackson13}.

We explain in Appendix \ref{pd_2step} how the asymptotic distribution of such an estimator can be derived with incomplete data, as well as how realisations from this distribution can be obtained. Following such developments, imputation of variable $y$ is performed as follows:
\begin{enumerate}[label=Step (\arabic*)]
\item $\PI$ is drawn according to the asymptotic posterior (Equations \ref{fcs2step1}-\ref{fcs2step4} in Appendix \ref{pd_2step}).
\item $y^{\text{miss}}\vert Y^{\text{obs}}, \PI$ is generated by:
\begin{itemize}
\item drawing $b_{\nbclust}$ from $\mathcal{N}\left(0,\bfPsi\right)$ for $1\leq\nbclust\leq\Nbclust$ if $y_{\nbclust}$ is systematically missing or conditionally on $\widehat{\bfbeta}_\nbclust$ if $y_{\nbclust}$ is sporadically missing \label{fcs2step5}
\item drawing $y_{\nbclust\nbind}^{\text{miss}}$ from $\mathcal{N}\left(\bfz_{\nbclust\nbind}\left(\bfbeta+b_{\nbclust}\right),\sigma^2_{\nbclust}\right)$ for all $\nbclust$ $(1\leq\nbclust\leq\Nbclust)$ and for all $\nbind$ $(1\leq\nbind\leq\Nbind_{\nbclust})$ such that $y_{\nbclust\nbind}$, the observation $\nbind$ in cluster $\nbclust$, is missing.\label{fcs2step6}
\end{itemize}
\end{enumerate}

Originally, this method was dedicated to handle incomplete continuous variables only. We extend it to handle binary variables with both sporadically and systematically missing values by applying a logit link in the imputation model (\ref{model2step}). In this case, the two-stage estimator is based on logistic models at stage one. The missing values can be imputed according to a similar scheme as continuous variables.\\

Table \ref{tablesynth} sums up the main modelling assumptions of each MI method. In the next section, the consequences of such differences are highlighted in terms of inference from incomplete data.
\begin{table}[h]
\begin{center}
\caption{Synthesis of the modelling assumptions of the MI methods JM-jomo, FCS-GLM and FCS-2stage\label{tablesynth}}
\begin{tabular}{lp{2.5cm}p{2.5cm}p{2.5cm}}
\hline
&					heteroscedasticity assumption&link function for binary variables&strategy for proper MI\\
\hline 
JM-jomo&yes&probit&Bayesian modelling based on conjugate prior\\ \hline
FCS-GLM&no&logit&Bayesian modelling based on Jeffrey prior\\ \hline
FCS-2stage&	yes&logit&asymptotic method based on a two-stage estimator\\ \hline
\end{tabular}
\end{center}
\end{table}
\subsection{Properties}\label{properties}
\subsubsection{FCS or JM \label{fcsjm}}
Comparisons between FCS and JM methods have been extensively studied \citep{vb07, Lee10, Zhao09, Wagstaff11, Kropko14, Hughes14, Resche16, erler_dealing_2016}, particularly in settings without clustering. It is generally believed that FCS methods are less likely to yield biased imputations because they allow for more flexibility than JM methods. For multilevel data, the lack of flexibility for JM methods has been recently highlighted when the analysis model includes random slopes corresponding to incomplete variables \citep{Enders16}. However, FCS-methods raised other issues like the variables selection for conditional models. Furthermore, the theoretical background of FCS is not well understood and constitutes a current topic of research \citep{Zhu15, Liu14, Bartlett15}. Indeed, contrary to a Gibbs sampler, convergence towards a joint posterior distribution cannot generally be proven \citep{Kropko14; Hughes14; VB12}.  Nevertheless, simulation show that it does not affect the quality of imputation without clustering \citep{vanBuuren06}. 
In addition, estimation of conditional distributions is more computationally intensive than the estimation of a joint distribution.
\subsubsection{One-stage or two-stage estimator}
The two FCS approaches use different estimators of the imputation model: the FCS-GLM method uses the one-stage estimator of parameters of model (\ref{glm}), while the two-stage estimator uses the re-written model (\ref{model2step}). The one-stage estimator has the drawback to be computationally intensive and slow to converge \citep{Yucel02}, particularly with binary variables \citep{Noh07}. The two-stage estimator solves this computational time issue, but tends to have a larger variance \citep{Mathew10} and requires large clusters with binary outcome to avoid separability problems \citep{Albert84} and to reduce the small-sample bias of the ML estimator \citep{Firth93}. Furthermore, by using a limited number of observations at stage one, the FCS-2stage method is more prone to suffer of overfitting if the number of covariates or the number of missing values is large.
\subsubsection{Heteroscedasticity}
JM-jomo and FCS-2stage allow for heteroscedasticity of the imputation model, whereas FCS-GLM assumes homoscedastic error variances. It has previously been demonstrated that data generated from a joint homoscedastic model (similar to model~(\ref{glmmulti})) will yield heteroscedastic conditional distributions \citep{Resche16}. As a result, imputation models allowing for heteroscedasticity tend to yield more reliable imputations. Previous simulation studies seem to support this point \citep{vanBuuren10,Resche16}. However, homoscedasticity can be relevant when studies are very small, since it overcomes overfitting issues by shrinking cluster-specific parameter estimates towards their weighted average.
\subsubsection{Bayesian modelling or asymptotic strategy for \ref{MIstep1}}
JM-jomo and FCS-GLM consider an explicit Bayesian specification of the imputation model, which implies that uncertainty of $\PI$ is fully propagated. Conversely, FCS-2stage only propagates the asymptotic uncertainty, and may therefore be problematic in small samples. Regardless, in large samples, both approaches should yield similar results \citep[p.~216]{Little02}.

As a direct result of Bayesian modelling, JM-jomo and FCS-GLM require the specification of a prior distribution for $\PI$. Various priors have been proposed for hierarchical models such as model~(\ref{glm}) \citep[p.~456-506]{Robert07}. In general, it is recommended to use proper prior distributions when working with multivariate linear mixed effect models \citep{Yucel02}, as this helps to avoid convergence issues of the Gibbs sampler. To this purpose, JM-jomo considers conjugate prior distributions. A major advantage of using conjugate prior distributions is that drawing from the posterior distribution avoids the systematic recourse to MCMC methods. In particular, when using univariate linear mixed effect model with fully observed covariates, the posterior distribution becomes analytically tractable.

In contrast to JM-jomo, FCS-GLM considers the Jeffreys prior for drawing imputations. This prior is derived from the sampling distribution and can therefore be regarded as non-informative. The counterpart of using Jeffreys prior in GLMM is that the (joint) posterior distribution for $\PI$ is generally improper \citep{Natarajan00} even if each marginal distribution is proper. Note that FCS-GLM overcomes this potential issue by assuming independent posterior distributions for each components of $\PI$.

In conclusion, all methods are likely to yield different posterior distributions, particularly in the presence of small sample sizes.

\subsubsection{Binary variables\label{propbin}}
JM-jomo considers a probit link to model binary variables, while both FCS approaches consider a logit link. Although both link functions tend to yield similar predictions, the probit link is more convenient for imputation purposes in multilevel data. The underlying reason is that conditional distributions of random coefficients can be easily simulated with a probit link, because it is based on latent normal variables for which these conditional distributions are well known, but not for mixed models with a logistic link. As a result, imputation of sporadically missing values is achieved in the same way as systematically missing variables for FCS-GLM, i.e. by ignoring the relationship between the random effects and the observed values on the imputed variables. It implies that this method is not relevant with binary sporadically missing variables including few missing values. Conversely, for FCS-2stage it is still possible to draw random coefficients from the conditional distribution, by considering the distribution of the random coefficients conditionally on the ML estimates given at stage one. Nevertheless, because of the asymptotic unbiasedness property of the ML estimator for logistic regression models (used at stage one), performances of the FCS-2stage method for binary variables are deteriorated when all clusters contain few observed individuals. 

\section{Simulations}
\label{Simulation}
\subsection{Simulation design\label{simudesign}}
We consider a simulation study to assess the relative performance of the MI methods described in Section \ref{MImethods}. Based on aforementioned properties, we anticipate that FCS-GLM is problematic when imputing binary sporadically missing variables, that FCS-2stage is problematic in datasets with few participants and/or clusters, and that JM-jomo is sensitive to the proportion of missing values because of the influence of the prior distribution. For this reason, we here evaluate the proportion of systematically and sporadically missing values, the data type of imputed variables, the size of included clusters and the size of the total dataset. Other settings will be also investigated to cover a large range of practical cases. For all investigated configurations, we generated $\Nbsim=500$ complete datasets, after which we introduced missing values. Afterwards, we applied the MI methods on each incomplete dataset by considering $\Nbtab=5$ complete datasets, and obtained parameter estimates from the multiply imputed datasets.

\subsubsection{Data generation}
For each simulation, we generated a dataset with four variables $\left(y,x_1,x_2,x_3\right)$; $x_1$ and $x_3$ are continuous variables, and $x_2$ is a binary variable. The outcome variable $y$ (continuous or binary) is defined according to a GLMM where $x_1$ and $x_2$ are the covariates. We use $x_3$ as an auxiliary variable explaining the missing data mechanism. More precisely, data are simulated as follows:
\begin{enumerate}
\item Draw $\Nbclust$ realisations of the triplet of variables $\left(v_1, v_2, v_3\right)$ so that 
\begin{eqnarray}\label{model_latent}
\left(v_1, v_2,v_3\right)\sim \mathcal{N}\left({\bf{0}},
\bfSigma_{\bf{v}}\right)
\end{eqnarray}
where $\bfSigma_{\bf{v}}$ is a $\left(3\times 3\right)$ covariance matrix.
\item Draw two continuous variables $x_1$ and $x_3$ so that:
\begin{eqnarray}
\left(\begin{array}{c}x_{1\nbclust\nbind},x_{3\nbclust\nbind}\end{array}\right)\sim\mathcal{N}\left(\left(\begin{array}{c}\alpha_1+v_{1\nbclust},\alpha_3+v_{3\nbclust}\end{array}\right),\bfSigma_{\bf{x}}\right)
\label{model_marg_quanti}
\end{eqnarray}
where $\alpha_1$ and $\alpha_3$ are the fixed intercepts, $v_{1\nbclust}$ and $v_{3\nbclust}$ are the random intercepts for cluster $\nbclust$ drawn at previous step and $\bfSigma_{\bf{x}}$ is a $\left(2\times 2\right)$ covariance matrix.
\item Draw a binary variable $x_2$ according to the model:
\begin{eqnarray}\label{model_marg}
\mathrm{logit}\left(P(x_{2\nbclust \nbind}=1)\right)=\alpha_2+v_{2\nbclust}
\end{eqnarray}
where $\alpha_2$ is a fixed intercept and $v_{2\nbclust}$ is the random intercept for cluster $\nbclust$, drawn from (\ref{model_latent})
\item Draw a response variable $y$:
\begin{itemize}
\item for a continuous variable $y$:\end{itemize}
 \begin{eqnarray}
 y_{\nbclust \nbind}=\beta^{(0)}+\beta^{(1)}x_{1\nbclust \nbind}+\beta^{(2)}x_{2\nbclust \nbind}+u_{\nbclust}^{(0)}+u^{(1)}_{\nbclust}x_{1\nbclust \nbind}+\varepsilon_{\nbclust \nbind} \label{analysiscont}
\end{eqnarray}
\begin{list}{}{}
\item where $\varepsilon_{\nbclust \nbind}\sim \mathcal{N}\left(0,\sigma_y^2\right)$  and  $\left(u_{\nbclust}^{(0)},u_{\nbclust}^{(1)}\right)$ is the random effects for cluster $\nbclust$ so that $\left(u_{\nbclust}^{(0)},u_{\nbclust}^{(1)}\right)\sim\mathcal{N}\left(0,\Psi\right)$ with $\Psi=\left(\begin{array}{cc}
\psi_{00}&\psi_{01}\\
\psi_{01}&\psi_{11}
\end{array}\right)$
\end{list}
\begin{itemize}
\item for a binary variable $y$:
\end{itemize}
\begin{eqnarray}
~~\mathrm{logit}\left(P\left(y_{\nbclust \nbind}=1\right)\right)=\beta^{(0)}+\beta^{(1)}x_{1\nbclust \nbind}+\beta^{(2)}x_{2\nbclust \nbind}+u_{\nbclust}^{(0)}+u_{\nbclust}^{(1)}x_{1\nbclust \nbind} 
\label{analysisbin}
\end{eqnarray}
\begin{list}{}{}
\item with the same assumption for $\left(u_{\nbclust}^{(0)},u_{\nbclust}^{(1)}\right)$.
\end{list}
\end{enumerate}
Parameters are tuned to mimic the structure of GREAT data. This data set is an individual patient data meta-analysis provided by the GREAT Network \citep{great} which explores risk factors associated with short-term mortality in acute heart failure. It consists of 28 observational cohorts gathering characteristics, potential risk factors and outcomes of 11685 patients, corresponding respectively to the second and the first level of a multilevel structure. 
One challenge consists in explaining the left ventricular ejection fraction (LVEF), which is realised with an ultrasound, from biomarkers easier to measure, such as BNP, which is a blood biomarker, and the atrial fibrillation (AFIB). More detail on GREAT data are given in Appendix \ref{AnnexGREAT}.

Data on the variable BNP is used to generate the continuous covariates $x_1$ and $x_3$, while the variable AFIB is used to generate the binary covariate $x_2$. We used complete-case analysis to estimate parameters of the analysis model (\ref{analysiscont}). Conversely, we tuned the covariate distribution according to the posterior distribution estimated by the fully Bayesian approach (Section \ref{jomomethod}) implemented in the R package \textit{jomo} \citep{jomopackage}. Thus, unless otherwise specified, default parameters are specified as follows: $\Nbclust=28, 18\leq\Nbind_{\nbclust}\leq 1093$,

$\bfSigma_{\bf{v}}=\left(\begin{array}{ccc}
0.12&0.001&0.001\\
0.001&0.12&0.001\\
0.001&0.001&0.12
\end{array}\right)$, $\left(\alpha_1,\alpha_3\right)=\left(2.9,2.9\right)$, $\bfSigma_{\bf{x}}=\left(\begin{array}{cc}
0.36&0.108\\
0.108&0.36\\
\end{array}\right)$, $\alpha_2=0.42$, $\left(\beta^{(0)},\beta^{(1)},\beta^{(2)}\right)=\left(0.72,-0.11,0.03\right)$, $\bfPsi=\left(\begin{array}{cc}
0.0077& -0.0015\\
 -0.0015&0.0004
\end{array}\right)$, $\sigma_Y=0.15$.

They define the base-case configuration. Then, these parameters will be varied one-by-one to investigate more specific configurations that can affect the methods. Details about the parameters used for each case is provided in Appendix \ref{AnnexDesign} in Table \ref{summarysimu}.
\subsubsection{Missing data mechanisms}
Variables are independently systematically missing on $(x_1,x_2)$ with probability $\pi_{sys}$. In addition, for any clusters where each covariate is not systematically missing,
sporadically missing values are generated with probability $\pi_{spor}$. Unless otherwise specified $\pi_{sys}=0.25$ and $\pi_{spor}=0.25$. Thus, the proportion of missing values on $x_1$ and $x_2$ is roughly 0.44. Two missing data mechanisms are considered: a MCAR and a MAR. For the MCAR mechanism, sporadically missing data is generated independently to the data, while for the MAR mechanism, sporadically missing values are added 
according to the observed values of the auxiliary variable $x_3$. In both cases systematically missing values remain MCAR.
\subsubsection{Methods\label{methods_design}}
The simulation study evaluates a total of 10 methods. The reference methods are as follows:
\begin{itemize}
\item Full - Analysis of original dataset, before introduction of missing values
\item CC - Case-wise deletion of individuals with incomplete data
\end{itemize}
Further, we consider three methods that allow imputation of sporadically and systematically missing data in multilevel data by adopting random effects distributions. The performance of these methods is of primary interest in the current simulation study:
\begin{itemize}
\item JM-jomo \citep{Quartagno15}
\item FCS-GLM \citep{Jolani15}
\item FCS-2stage \citep{Resche16} (estimation using REML and MM)
\end{itemize}
Finally, we consider five ad-hoc methods that were not designed to be used in multilevel data with a combination of sporadically and systematically missing values. Nevertheless, these methods are evaluated to highlight the relative merits of aforementioned, dedicated, methods:
\begin{itemize}
\item JM-pan \citep{Yucel02}: JM imputation by linear mixed effect models assuming homoscedasticity
\item FCS-2lnorm \citep{vanBuuren10}: FCS imputation by linear mixed effect models assuming heteroscedasticity
\item FCS-noclust \citep{Schafer97}: FCS imputation by normal or logistic regression
\item FCS-fixclust: FCS imputation by normal or logistic regression with fixed intercept to account for the second-level
\item FCS-fixclustPMM \citep{Little88}: FCS imputation by predictive mean matching with fixed intercept to account for the second-level
\end{itemize}
As explained in the introduction, these later methods are not dedicated to multilevel data with continuous and binary variables, as well as sporadically and systematically missing values. JM-pan and FCS-2lnorm only allow imputation of continuous data (Section \ref{MImethod}). For this reason, binary variables are treated as continuous, without applying any rounding strategy \citep{Allison02}. Furthermore, because of the heteroscedasticity assumption, parameters of FCS-2lnorm are not identifiable in the presence of systematically missing values. We address the issue by imputing sporadically missing values and systematically missing values separately from each other. In particular, clusters without systematically missing data are used to fit the imputation model and to impute clusters with sporadically missing values. Afterwards, parameters obtained from the first clusters without systematically missing data are used to impute the remaining clusters with systematically missing data. As regard FCS-noclust, FCS-fixclust and FCS-fixclustPMM, these methods are based on fixed intercept, implying non-identifiability of the intercept with systematically missing variables. The issue is addressed by centring dummy variables such that clusters with systematically missing values are imputed using the observed average across the remaining clusters.
\subsubsection{Assessment of the inference}
The primary parameters of interest are $\beta^{(1)}$, $\beta^{(2)}$, $\psi_{00}$ and $\psi_{11}$ in model (\ref{analysiscont}) or (\ref{analysisbin}). The performance of estimating these parameters will be assessed according to the bias, the root mean squared error (RMSE), the root mean square of estimated standard error (Model SE), the empirical Monte Carlo standard error (Emp SE) and the coverage of the associated confidence interval. The average time required to multiply impute one dataset is also reported.

\subsubsection{Implementation\label{implementation}}
Simulations are performed with the R software \citep{Rsoft}. Multiple imputation with the JM-jomo method is performed with the R package \textit{jomo} \citep{jomopackage}. The number of iterations for the burn-in step is set to 2000, and 1000 iterations are run between each imputed dataset. Convergence has been checked from an incomplete dataset simulated from the base-case configuration by following the stationary of the parameters of the imputation model. 

Multiple imputation with FCS methods is performed using the R package \textit{mice} \citep{mice} with 5 cycles. Convergence has been checked from an incomplete dataset simulated from the base-case configuration by following the stationary of marginal quantities (means and standard deviations). For both FCS approaches, conditional imputation models consist of all available covariates, which are included in the fixed and random design matrices ($\bf Z_k=\bf W_k$).

In both cases, MI is performed using 5 imputed datasets. Each imputed dataset is analysed using the R package \textit{nlme} \citep{nlme} for a continuous outcome and using the \textit{glmer} package for a binary outcome \citep{lme4}. Calculation was performed on an Intel\textregistered  Xeon\textregistered CPU E7530 1.87GHz.
The R code used to perform simulations is available from \texttt{https://arxiv.org/src/1702.00971v1/anc/}
 
\subsection{Results}
\subsubsection{Base-case configuration\label{BCconfig}}
Table \ref{tablebasecase} describes the simulation study results for the base-case configuration. Overall, we found that all methods yielded satisfactory estimates for fixed effects parameters of the binary variable $\beta^{(2)}$. In particular, biases were smaller than 2\%, and coverage of the confidence intervals were close to their nominal level. Performance differences mainly occurred for estimation of fixed effects parameters for the continuous variable $\beta^{(1)}$ and for estimation of the variance of random effects $\psi_{00}$ and $\psi_{11}$. In particular, the variance of $\widehat{\beta^{(1)}}$ is generally underestimated, leading to confidence intervals that do not reach their nominal level.

As expected, we found that ad-hoc methods suffered from several deficiencies. First of all, they underestimated the variance of $\widehat{\beta^{(1)}}$. This is likely related to the use of imputation models with homoscedastic error terms (FCS-noclust, FCS-fixclust, FCS-fixclustPMM and JM-pan), and to not properly modelling heterogeneity between clusters (FCS-noclust, FCS-fixclust and FCS-fixclustPMM). For FCS-2lnorm, underestimation of the variance is likely also caused by ignoring uncertainty on the random coefficients for systematically missing values. A second problem of the ad-hoc methods is that they yield severely biased estimates for the variance of random effects ($\psi_{00}$ and $\psi_{11}$). Finally, FCS-noclust also introduces bias on point estimates for the fixed effects coefficients. In particular, by ignoring the multilevel data structure, imputed values of FCS-noclust are biased towards the overall mean and thereby affect corresponding covariate-outcome associations.

The primary methods of interest, JM-jomo, FCS-GLM and FCS-2stage provide inferences that are more satisfying as compared to the ad-hoc methods. In particular,  biases are smaller and confidence intervals are closer to their nominal level. Nevertheless, some important differences were identified. First of all, the JM-jomo method tends to overestimate the variance of the estimators $\widehat{\beta^{(1)}}$ and $\widehat{\beta^{(2)}}$. Conversely, FCS-GLM tends to underestimate this variance for $\widehat{\beta^{(1)}}$, similar to ad-hoc methods assuming homoscedasticity. Another drawback of FCS-GLM is that its implementation required substantially more time to generate an imputed dataset. Finally, the FCS-2stage method provided satisfactory inferences with both versions (REML or MM). In particular, it is the only method to provide unbiased estimates for variance components $\psi_{00}$ and $\psi_{11}$.

\begin{sidewaystable}
\centering
\caption{Simulation study results from the base-case configuration. Point estimate, relative bias, model standard error, empirical standard error, 95\% coverage and RMSE for analysis model's parameters and for several methods (Full data, Complete-case analysis, FCS-noclust, FCS-fixclust, FCS-fixclustPMM, JM-pan, FCS-2lnorm, JM-jomo , FCS-GLM, FCS-2stage with REML estimator, FCS-2stage with moment estimator). Criteria are based on 500 incomplete datasets. Average time to multiply impute one dataset is indicated in minutes. Criteria related to the continuous (resp. binary) covariate are in light (resp. dark) grey. True values are $\beta^{(1)}=$-0.11, $\beta^{(2)}=$0.03, $\sqrt{\psi_{00}}=0.088$, $\sqrt{\psi_{11}}=0.02$. \label{tablebasecase}} 
\begingroup\footnotesize
\begin{tabular}{lrrrrrrrrrrr}
  \hline
\begin{sideways}   \end{sideways} & \begin{sideways} Full \end{sideways} & \begin{sideways} CC \end{sideways} & \begin{sideways} FCS-noclust \end{sideways} & \begin{sideways} FCS-fixclust \end{sideways} & \begin{sideways} FCS-fixclustPMM \end{sideways} & \begin{sideways} JM-pan \end{sideways} & \begin{sideways} FCS-2lnorm \end{sideways} & \begin{sideways} JM-jomo \end{sideways} & \begin{sideways} FCS-GLM \end{sideways} & \begin{sideways} FCS-2stageREML \end{sideways} & \begin{sideways} FCS-2stageMM \end{sideways} \\ 
  \rowcolor[gray]{.9}  \hline
$\beta^{(1)}$ est & -0.1101 & -0.1104 & -0.1102 & -0.1102 & -0.1039 & -0.1088 & -0.1098 & -0.1087 & -0.1100 & -0.1089 & -0.1090 \\ 
   \rowcolor[gray]{.9} $\beta^{(1)}$ rbias (\%) & 0.1 & 0.3 & 0.2 & 0.2 & -5.5 & -1.1 & -0.2 & -1.2 & -0.0 & -1.0 & -0.9 \\ 
   \rowcolor[gray]{.9} $\beta^{(1)}$ model se & 0.0047 & 0.0070 & 0.0043 & 0.0043 & 0.0042 & 0.0044 & 0.0056 & 0.0068 & 0.0047 & 0.0059 & 0.0059 \\ 
   \rowcolor[gray]{.9} $\beta^{(1)}$ emp se & 0.0048 & 0.0071 & 0.0057 & 0.0058 & 0.0066 & 0.0056 & 0.0063 & 0.0057 & 0.0057 & 0.0058 & 0.0058 \\ 
   \rowcolor[gray]{.9} $\beta^{(1)}$ 95\% cover & 93.8 & 92.2 & 86.0 & 85.6 & 60.4 & 87.6 & 92.2 & 97.6 & 91.1 & 95.0 & 95.0 \\ 
   \rowcolor[gray]{.9} $\beta^{(1)}$ rmse & 0.0048 & 0.0071 & 0.0057 & 0.0058 & 0.0090 & 0.0058 & 0.0063 & 0.0058 & 0.0057 & 0.0059 & 0.0059 \\ 
   \rowcolor[gray]{.7}  \hline
$\beta^{(2)}$ est & 0.0301 & 0.0299 & 0.0300 & 0.0300 & 0.0290 & 0.0295 & 0.0301 & 0.0297 & 0.0297 & 0.0295 & 0.0297 \\ 
   \rowcolor[gray]{.7} $\beta^{(2)}$ rbias (\%) & 0.2 & -0.4 & 0.2 & -0.0 & -3.2 & -1.7 & 0.2 & -1.1 & -1.1 & -1.6 & -1.0 \\ 
   \rowcolor[gray]{.7} $\beta^{(2)}$ model se & 0.0029 & 0.0053 & 0.0044 & 0.0043 & 0.0043 & 0.0044 & 0.0064 & 0.0069 & 0.0046 & 0.0054 & 0.0049 \\ 
   \rowcolor[gray]{.7} $\beta^{(2)}$ emp se & 0.0030 & 0.0053 & 0.0043 & 0.0042 & 0.0044 & 0.0042 & 0.0055 & 0.0049 & 0.0042 & 0.0044 & 0.0044 \\ 
   \rowcolor[gray]{.7} $\beta^{(2)}$ 95\% cover & 94.2 & 94.4 & 95.2 & 95.4 & 93.6 & 95.6 & 95.4 & 97.2 & 94.2 & 97.0 & 96.2 \\ 
   \rowcolor[gray]{.7} $\beta^{(2)}$ rmse & 0.0030 & 0.0053 & 0.0043 & 0.0042 & 0.0045 & 0.0042 & 0.0055 & 0.0049 & 0.0043 & 0.0045 & 0.0044 \\ 
   \hline
$\sqrt{\psi_0}$ est & 0.0859 & 0.0835 & 0.0747 & 0.0745 & 0.0712 & 0.0806 & 0.0816 & 0.0941 & 0.0783 & 0.0869 & 0.0863 \\ 
  $\sqrt{\psi_0}$ rbias (\%) & -2.1 & -4.8 & -14.9 & -15.1 & -18.8 & -8.2 & -7.0 & 7.3 & -10.8 & -0.9 & -1.6 \\ 
  $\sqrt{\psi_0}$ rmse & 0.0154 & 0.0240 & 0.0189 & 0.0191 & 0.0213 & 0.0146 & 0.0166 & 0.0150 & 0.0171 & 0.0148 & 0.0150 \\ 
   \rowcolor[gray]{.9}  \hline
$\sqrt{\psi_1}$ est & 0.0193 & 0.0184 & 0.0138 & 0.0138 & 0.0135 & 0.0137 & 0.0174 & 0.0224 & 0.0152 & 0.0197 & 0.0194 \\ 
   \rowcolor[gray]{.9} $\sqrt{\psi_1}$ rbias (\%) & -3.7 & -8.1 & -30.9 & -31.1 & -32.3 & -31.4 & -13.2 & 12.0 & -24.2 & -1.3 & -2.9 \\ 
   \rowcolor[gray]{.9} $\sqrt{\psi_1}$ rmse & 0.0041 & 0.0073 & 0.0073 & 0.0074 & 0.0075 & 0.0074 & 0.0053 & 0.0043 & 0.0066 & 0.0046 & 0.0048 \\ 
   \hline
time &  &  & 1.8 & 1.2 & 0.9 & 1.4 & 3.3 & 8.0 & 114.7 & 2.5 & 1.0 \\ 
   \hline
\end{tabular}
\endgroup
\end{sidewaystable}

\subsubsection{Robustness to the proportion of systematically missing values}
To assess the influence of the proportion of systematically missing values, we modified this proportion to 0.1, 0.25 and 0.4 (configurations 2, 1, 3 in Table \ref{summarysimu} in Appendix \ref{AnnexDesign} respectively). The proportion of sporadically missing values is modified accordingly to keep the same proportion of missing values in expectation.

We found that for the three methods of primary interest, the bias on the parameters of the model remains stable regardless the proportion of systematically missing values (see Appendix \ref{AnnexSyst}). Figure \ref{fig7} reports the relative bias for the standard error estimates. For JM-jomo, the standard error estimate for $\beta^{(1)}$ tends to deviate from the empirical standard error when the proportion of systematically missing values increases, becoming upwardly biased as the relative extent of systematically missing data increases. This issue is likely related to the use of (informative) conjugate prior distributions, as their influence on the posterior is substantial when the proportion of systematically missing variables is large. Because the FCS methods use prior distributions that are derived from the data, they were less sensitive to overestimation of standard errors.

\begin{figure}
\begin{center}
\caption{Robustness to the proportion of systematically missing values: estimate of the relative bias for the SE estimate for $\widehat{\beta^{(1)}}$ (left), $\widehat{\beta^{(2)}}$ (right) according to $\pi_{syst}$ for each MI method. The estimated relative bias is calculated by the difference between the model SE and the empirical SE, divided by the empirical SE. The proportion of sporadically missing values is accordingly modified to keep a constant proportion of missing values (in expectation).\label{fig7}}
\includegraphics[width=6cm, height=8cm]{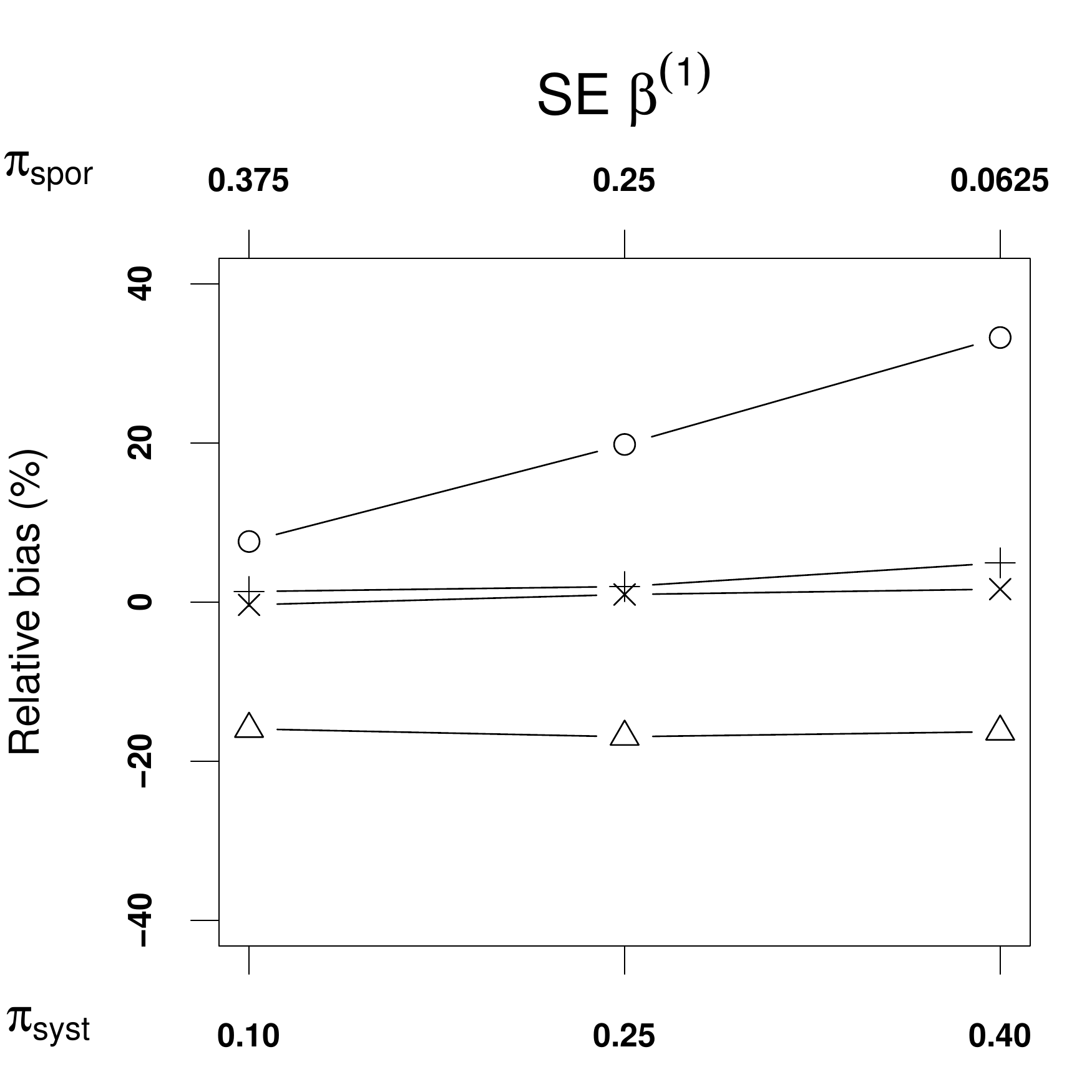} \includegraphics[width=6cm, height=8cm]{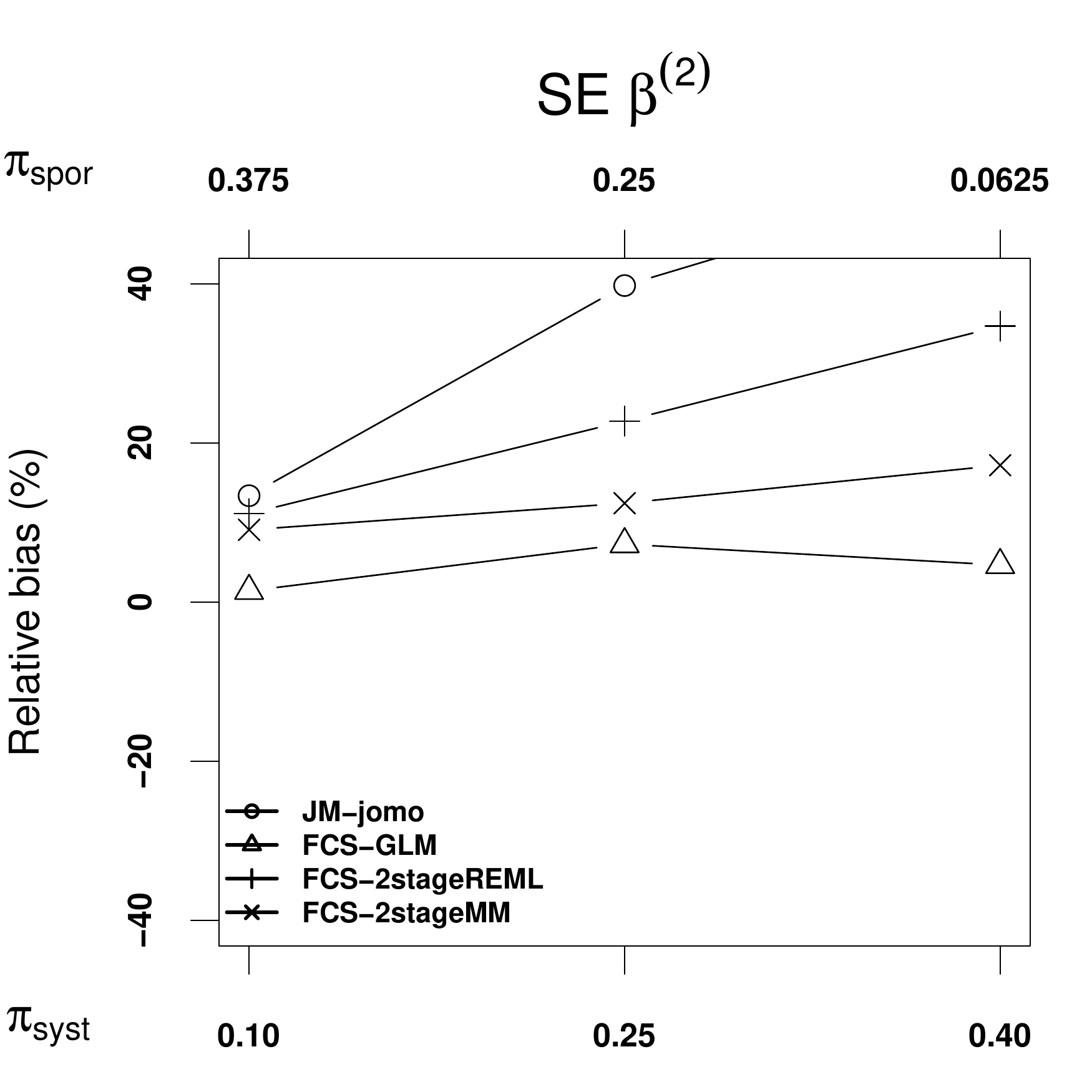}
\end{center}
\end{figure}
\subsubsection{Robustness to the number of clusters}
\begin{figure}
\begin{center}
\caption{Robustness to the number of clusters: relative bias for the estimate of $\beta^{(1)}$ (left), $\beta^{(2)}$ (middle) and $\psi_{11}$ (right) according to $\Nbclust$ for each MI method.\label{fig8}}
\includegraphics[width=4cm, height=6cm]{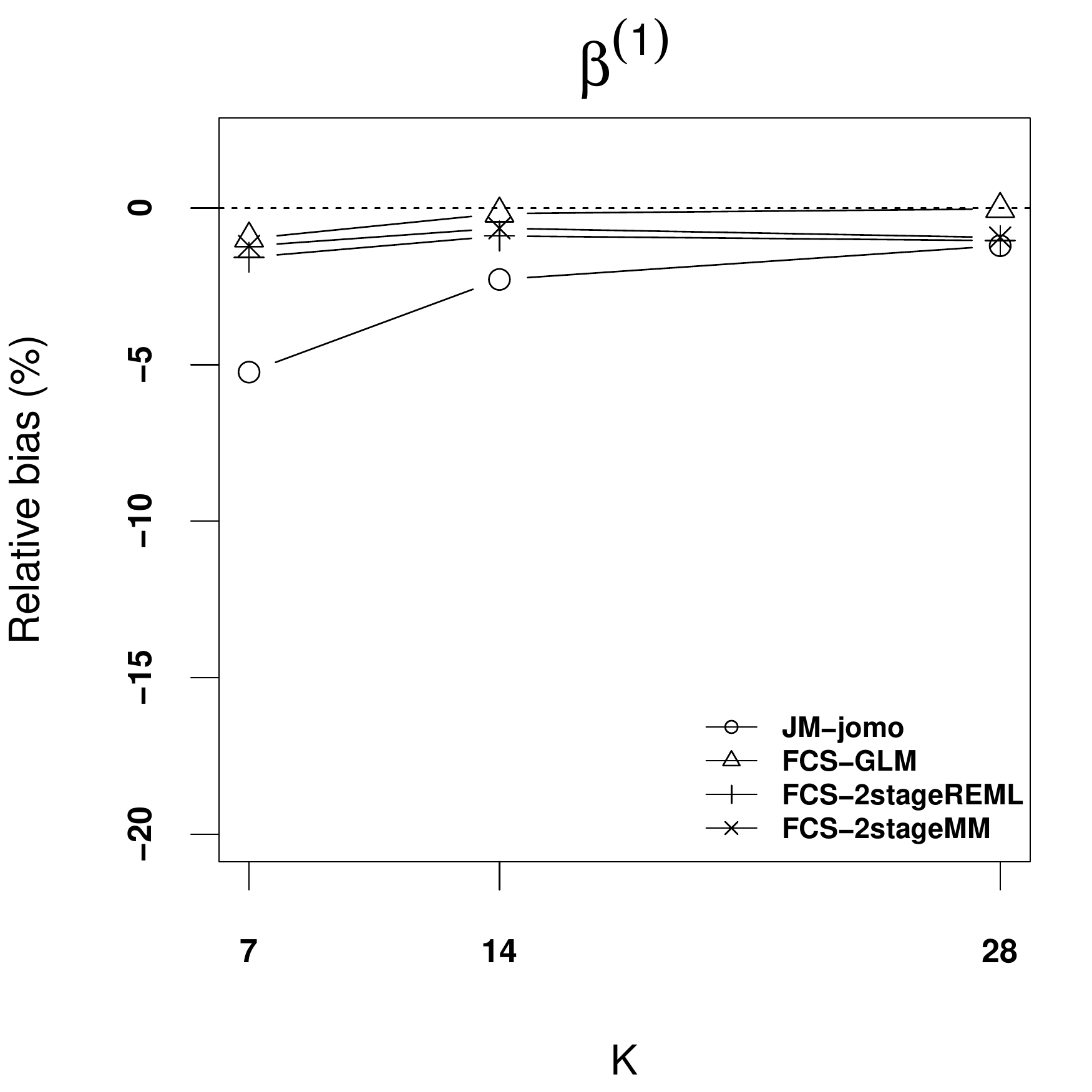} \includegraphics[width=4cm, height=6cm]{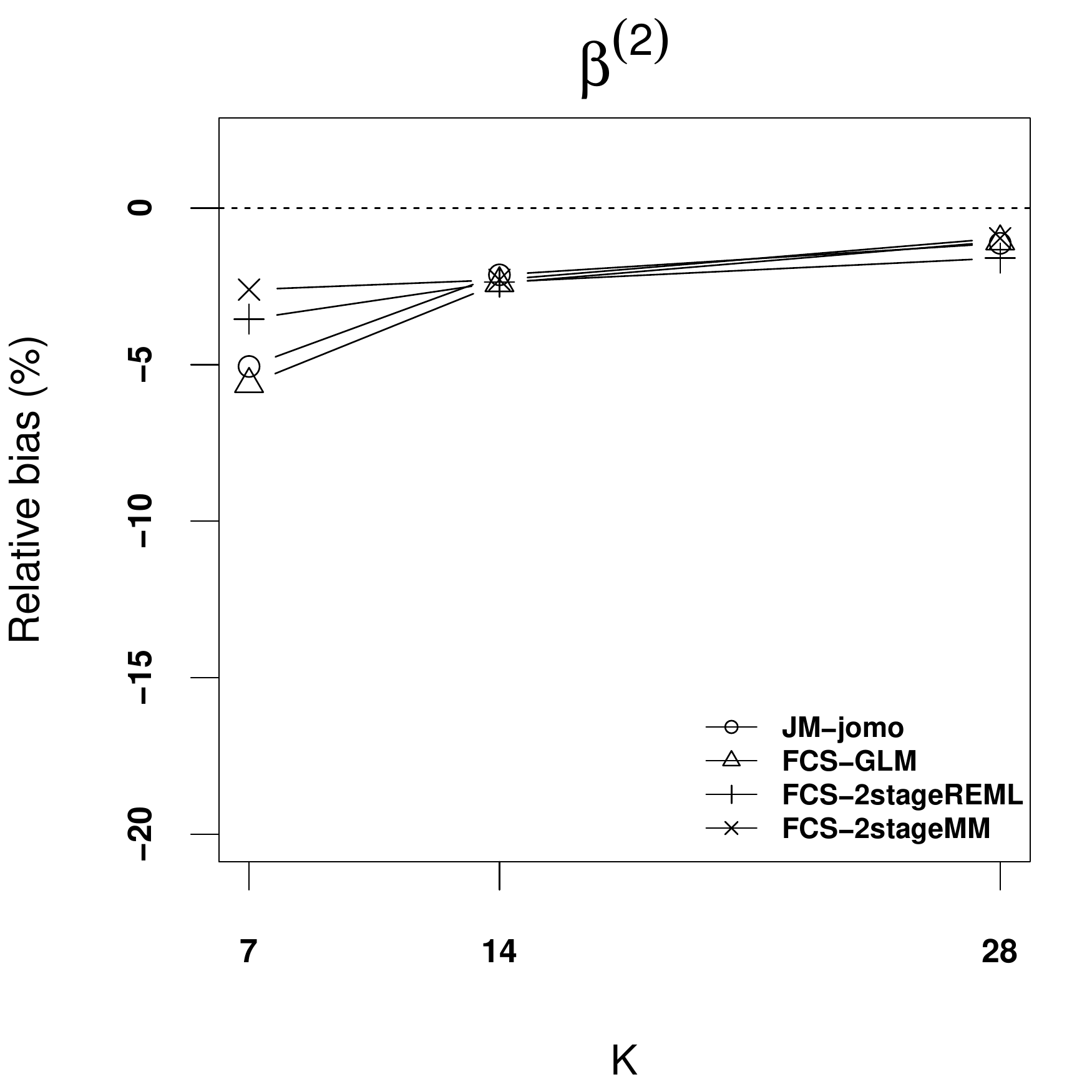}\includegraphics[width=4cm, height=6cm]{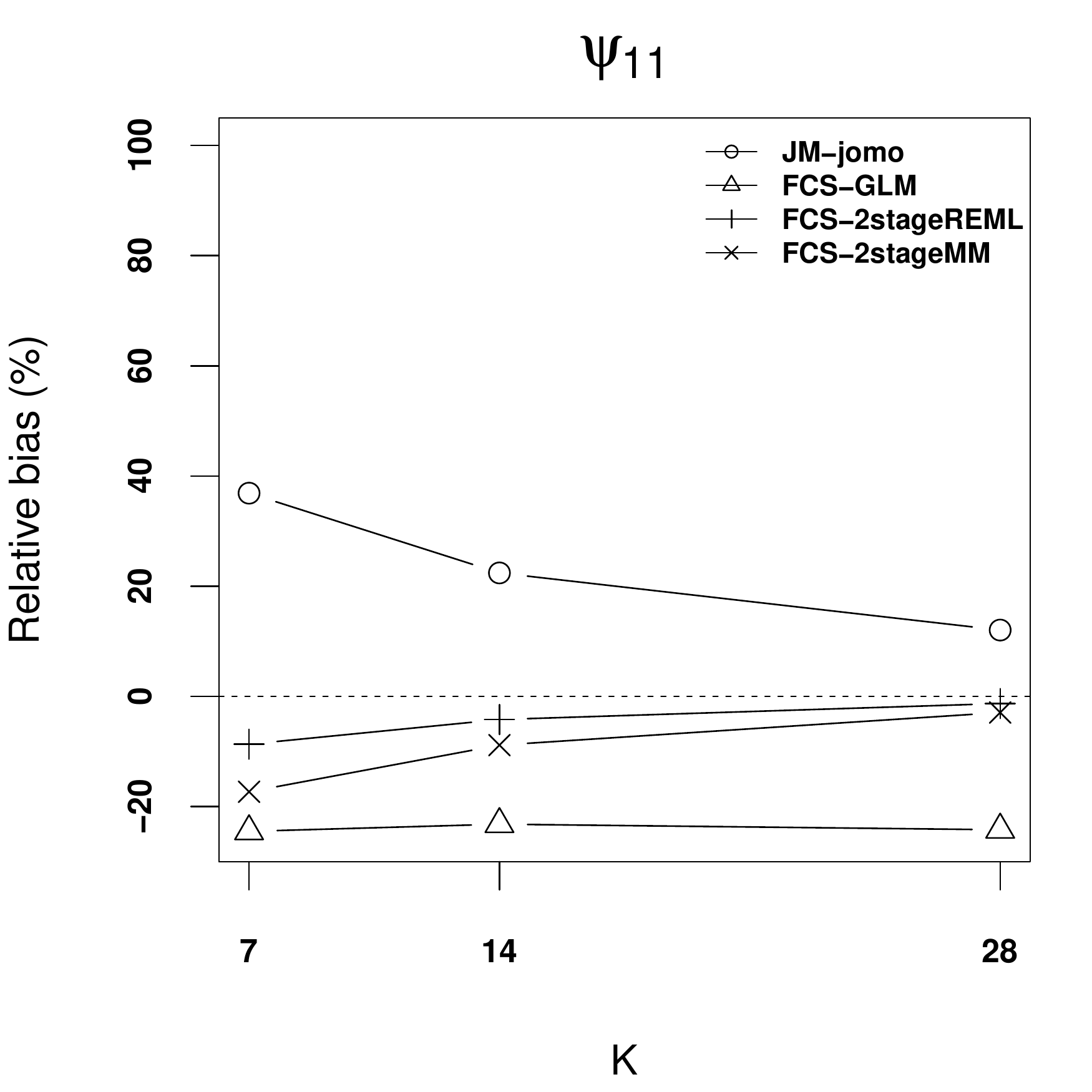}
\end{center}
\end{figure}

Influence of the number of clusters is assessed by restricting the generated datasets to their $\Nbclust$ first clusters, and varying $\Nbclust$ in $\lbrace 7, 14, 28 \rbrace$. Note that as consequence the total sample size also increases with $\Nbclust$ (2139, 4256 and  11685 respectively). Figure \ref{fig8} describes the impact of the number of clusters on the bias. The impact on the variance estimate is reported in Appendix \ref{AnnexK}.

For all estimands, the bias obtained from the JM-jomo method is substantial when the number of clusters is small, but decreases when as the number of clusters increases. On the contrary, the FCS methods provide more robust estimates for the variance of the random effects when the number of clusters is small. This behaviour is again likely related to the choice of the prior distributions.
\subsubsection{Robustness to the cluster size\label{robclustsize}}
\begin{figure}
\begin{center}
\caption{Robustness to the cluster size: relative bias for the estimate of $\beta^{(1)}$ (left), $\beta^{(2)}$ (middle) and $\psi_{11}$ (right) according to $\Nbind_{\nbclust}$ for each MI method. Criteria are based on 500 incomplete datasets.\label{fig4}}
\includegraphics[width=4cm, height=6cm]{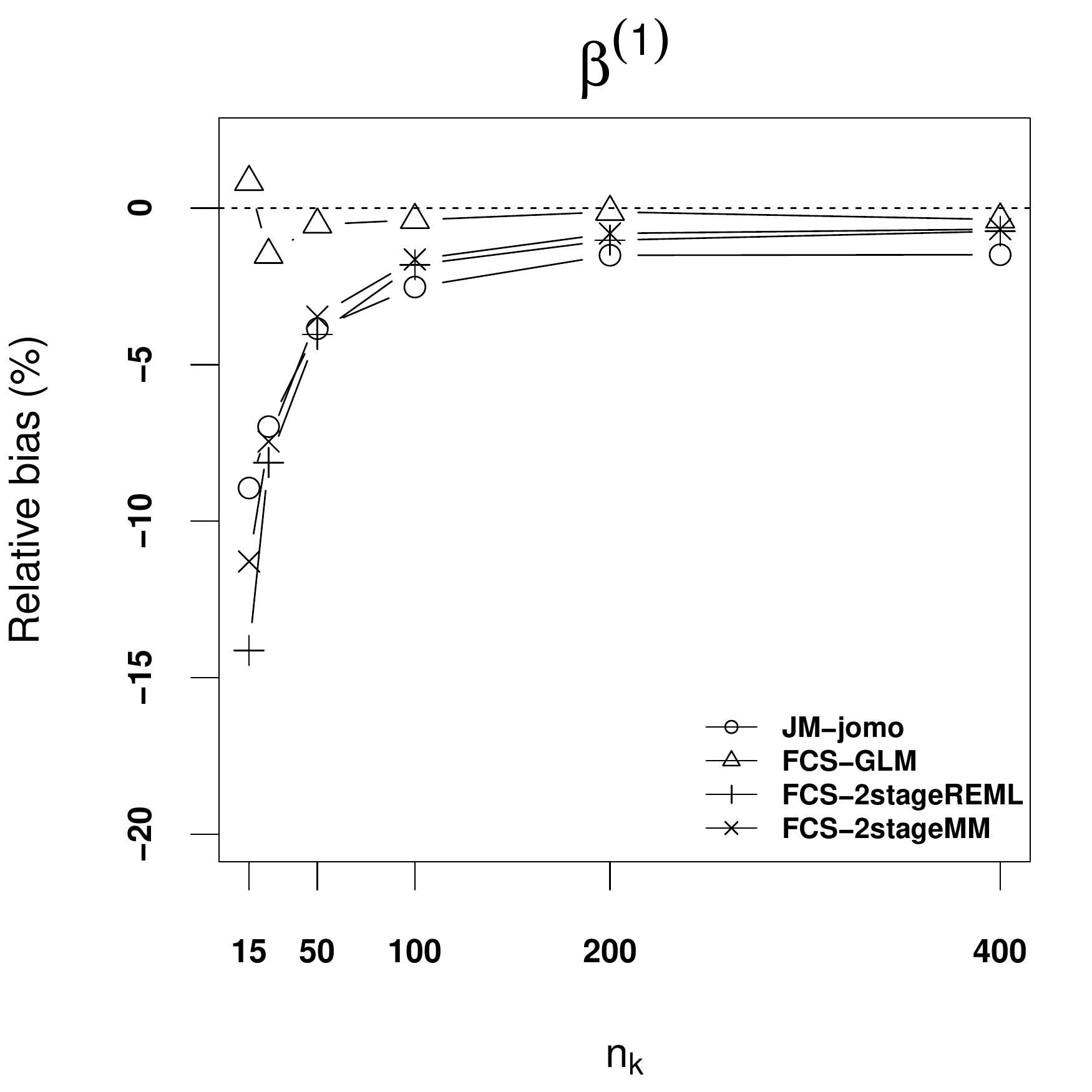} \includegraphics[width=4cm, height=6cm]{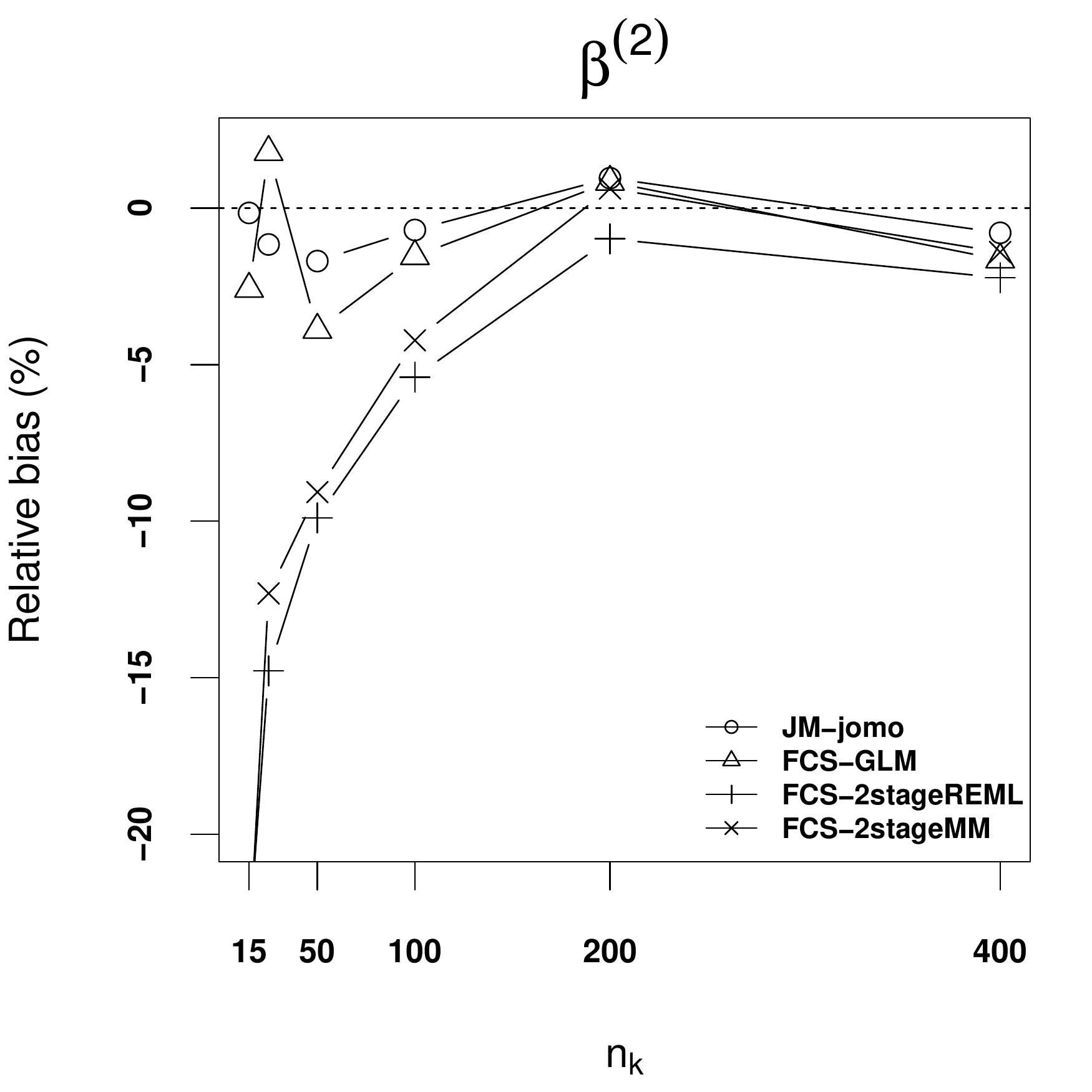}\includegraphics[width=4cm, height=6cm]{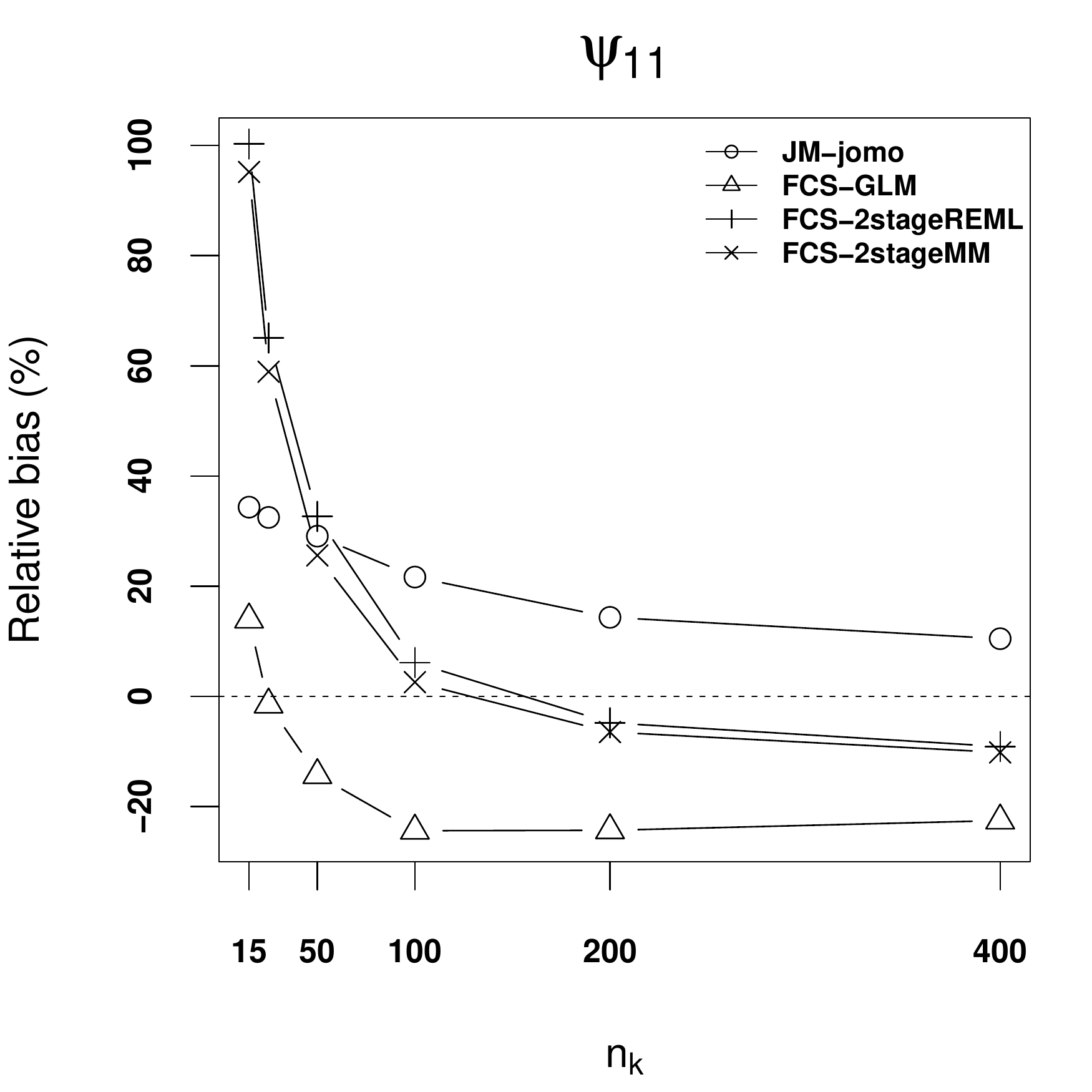}
\end{center}
\end{figure}

To explore the robustness of the inferences provided by the MI methods with respect to the size of the clusters, we extended the simulation study to generate clusters with equal sizes varying in $\lbrace 15, 25, 50, 100, 200, 400\rbrace$. Relative biases are reported in Figure \ref{fig4}.

The biases obtained by the FCS-2stage methods are large for small clusters, but decrease when the cluster size increases. This behaviour was expected since the posterior distribution for the conditional imputation models's parameters are based on asymptotic properties (\textit{cf} Section \ref{properties}). For the JM-jomo method, the bias mainly depends on the sample size for $\beta^{(1)}$ only. On the contrary, the FCS-GLM method is fairly stable for $\beta^{(1)}$ and $\beta^{(2)}$ across the sample size. Note that the bias on $\psi_{11}$ observed on the base-case configuration disappears with small clusters as well as the undercoverage issue because of a better estimate of the standard error (see Figure \ref{figa3} in appendix \ref{Annexnk}) making the method relevant in such a case.

\subsubsection{Robustness to the type of imputed variables}
Results in Table \ref{config4} demonstrate that also the type of imputed variable affects the performance of the imputation methods. In general, we found that JM-jomo provides smaller bias for $\beta^{(1)}, \beta^{(2)}$ and $\psi_{11}$ as compared to  FCS-GLM and FCS-2stage. This is likely related to the fact that FCS-GLM is not tailored for imputing sporadically missing data of binary variables, and that the two-stage estimator used in FCS-2stage is known to be biased in the presence of small clusters.
\begin{table}
\centering
\begingroup\footnotesize
\caption{Binary covariates. Point estimate, relative bias, model standard error, empirical standard error, 95\% coverage and RMSE for analysis model's parameters and for several methods (Full data, JM-jomo, FCS-GLM, FCS-2stage with REML estimator, FCS-2stage with moment estimator). Criteria are based on 500 incomplete datasets. Average time to multiply impute one dataset is indicated in minutes. True values are $\beta^{(1)}=$-0.11, $\beta^{(2)}=$0.03, $\sqrt{\psi_{00}}=0.088$, $\sqrt{\psi_{11}}=0.02$.\label{config4}} 
\begin{tabular}{lrrrrr}
  \hline
\begin{sideways}   \end{sideways} & \begin{sideways} Full \end{sideways} & \begin{sideways} JM-jomo \end{sideways} & \begin{sideways} FCS-GLM \end{sideways} & \begin{sideways} FCS-2stageREML \end{sideways} & \begin{sideways} FCS-2stageMM \end{sideways} \\ 
    \hline
$\beta^{(1)}$ est & -0.1098 & -0.1091 & -0.1081 & -0.1080 & -0.1083 \\ 
  $\beta^{(1)}$ rbias (\%) & -0.2 & -0.8 & -1.7 & -1.8 & -1.5 \\ 
  $\beta^{(1)}$ model se & 0.0050 & 0.0074 & 0.0057 & 0.0063 & 0.0056 \\ 
  $\beta^{(1)}$ emp se & 0.0049 & 0.0064 & 0.0059 & 0.0060 & 0.0061 \\ 
  $\beta^{(1)}$ 95\% cover & 95.0 & 97.0 & 92.0 & 94.0 & 90.4 \\ 
  $\beta^{(1)}$ rmse & 0.0049 & 0.0064 & 0.0062 & 0.0063 & 0.0063 \\ 
   \hline
$\beta^{(2)}$ est & 0.0303 & 0.0298 & 0.0295 & 0.0294 & 0.0295 \\ 
  $\beta^{(2)}$ rbias (\%) & 1.0 & -0.6 & -1.8 & -2.1 & -1.6 \\ 
  $\beta^{(2)}$ model se & 0.0029 & 0.0072 & 0.0044 & 0.0051 & 0.0045 \\ 
  $\beta^{(2)}$ emp se & 0.0028 & 0.0047 & 0.0043 & 0.0044 & 0.0043 \\ 
  $\beta^{(2)}$ 95\% cover & 95.0 & 98.6 & 95.2 & 96.2 & 95.0 \\ 
  $\beta^{(2)}$ rmse & 0.0028 & 0.0047 & 0.0044 & 0.0044 & 0.0044 \\ 
   \hline
$\sqrt{\psi_{00}}$ est & 0.0876 & 0.0861 & 0.0807 & 0.0834 & 0.0827 \\ 
  $\sqrt{\psi_{00}}$ rbias (\%) & -0.2 & -1.9 & -8.0 & -4.9 & -5.7 \\ 
  $\sqrt{\psi_{00}}$ rmse & 0.0123 & 0.0122 & 0.0137 & 0.0129 & 0.0131 \\ 
   \hline
$\sqrt{\psi_{11}}$ est & 0.0198 & 0.0232 & 0.0151 & 0.0185 & 0.0164 \\ 
  $\sqrt{\psi_{11}}$ rbias (\%) & -0.8 & 16.2 & -24.3 & -7.7 & -17.8 \\ 
  $\sqrt{\psi_{11}}$ rmse & 0.0046 & 0.0053 & 0.0069 & 0.0057 & 0.0068 \\ 
   \hline
time &  & 5.6 & 95.1 & 1.4 & 0.6 \\ 
   \hline
\end{tabular}
\endgroup
\end{table}
\subsubsection{Robustness to the variance of random effects}\label{robvarrand}
Table \ref{Tablenew} provides inference results when the covariance matrix of random effects is multiplied by a factor 2. The biases reported for the variance of random effects are less than 2\% for JM-jomo, while they reached 11\% in the base-case configuration. Such behaviour can be explained by the smaller influence of the prior distribution for random effects when the effect of random effects is stronger. On the contrary, the bias increases for the FCS-2stage methods.
\begin{table}[H]
\centering
\caption{Higher variance for $\bfPsi$. Point estimate, relative bias, model standard error, empirical standard error, 95\% coverage and RMSE for analysis model's parameters and for several methods (Full data, JM-jomo , FCS-GLM, FCS-2stage with REML estimator, FCS-2stage with moment estimator). Criteria are based on 500 incomplete datasets. Average time to multiply impute one dataset is indicated in minutes. Criteria related to the continuous (resp. binary) covariate are in light (resp. dark) grey. True values are $\beta^{(1)}=$-0.11, $\beta^{(2)}=$0.03, $\sqrt{\psi_{00}}=0.124$, $\sqrt{\psi_{11}}=0.028$.} 
\begingroup\footnotesize
\begin{tabular}{lrrrrr}
  \hline
& \sidewaysc{ Full } & \sidewaysc{ JM-jomo } & \sidewaysc{ FCS-GLM } & \sidewaysc{ FCS-2stageREML } & \sidewaysc{ FCS-2stageMM } \\ 
  \rowcolor[gray]{.9}  \hline
$\beta^{(1)}$ est & -0.1102 & -0.1087 & -0.1099 & -0.1089 & -0.1090 \\ 
   \rowcolor[gray]{.9} $\beta^{(1)}$ rbias (\%) & 0.2 & -1.1 & -0.1 & -1.0 & -0.9 \\ 
   \rowcolor[gray]{.9} $\beta^{(1)}$ model se & 0.0061 & 0.0075 & 0.0059 & 0.0070 & 0.0070 \\ 
   \rowcolor[gray]{.9} $\beta^{(1)}$ emp se & 0.0063 & 0.0071 & 0.0072 & 0.0073 & 0.0073 \\ 
   \rowcolor[gray]{.9} $\beta^{(1)}$ 95\% cover & 93.8 & 95.0 & 90.1 & 93.0 & 93.8 \\ 
   \rowcolor[gray]{.9} $\beta^{(1)}$ rmse & 0.0063 & 0.0073 & 0.0072 & 0.0073 & 0.0073 \\ 
   \rowcolor[gray]{.7}  \hline
$\beta^{(2)}$ est & 0.0301 & 0.0297 & 0.0295 & 0.0294 & 0.0296 \\ 
   \rowcolor[gray]{.7} $\beta^{(2)}$ rbias (\%) & 0.2 & -0.8 & -1.6 & -2.0 & -1.3 \\ 
   \rowcolor[gray]{.7} $\beta^{(2)}$ model se & 0.0029 & 0.0070 & 0.0046 & 0.0055 & 0.0050 \\ 
   \rowcolor[gray]{.7} $\beta^{(2)}$ emp se & 0.0030 & 0.0048 & 0.0042 & 0.0044 & 0.0044 \\ 
   \rowcolor[gray]{.7} $\beta^{(2)}$ 95\% cover & 94.2 & 98.2 & 94.2 & 97.4 & 96.0 \\ 
   \rowcolor[gray]{.7} $\beta^{(2)}$ rmse & 0.0030 & 0.0048 & 0.0043 & 0.0044 & 0.0044 \\ 
   \hline
$\sqrt{\psi_0}$ est & 0.1220 & 0.1225 & 0.1089 & 0.1172 & 0.1166 \\ 
  $\sqrt{\psi_0}$ rbias (\%) & -1.7 & -1.3 & -12.2 & -5.6 & -6.0 \\ 
  $\sqrt{\psi_0}$ rmse & 0.0198 & 0.0175 & 0.0240 & 0.0203 & 0.0205 \\ 
   \rowcolor[gray]{.9}  \hline
$\sqrt{\psi_1}$ est & 0.0275 & 0.0279 & 0.0220 & 0.0262 & 0.0260 \\ 
   \rowcolor[gray]{.9} $\sqrt{\psi_1}$ rbias (\%) & -2.8 & -1.5 & -22.1 & -7.3 & -8.2 \\ 
   \rowcolor[gray]{.9} $\sqrt{\psi_1}$ rmse & 0.0048 & 0.0043 & 0.0083 & 0.0057 & 0.0059 \\ 
   \hline
time &  & 7.7 & 102.5 & 2.2 & 0.9 \\ 
   \hline
\end{tabular}
\endgroup
\label{Tablenew}
\end{table}
\subsubsection{Other configurations}
Other configurations that have been investigated are presented in Appendix \ref{AnnexDesign}. These configurations consider the nature of the outcome (configuration 5), the missing data mechanism for sporadically missing values (configurations 6, 7, 8), the complexity of the analysis model (configuration 9), the number of individuals with unequal cluster sizes (configuration 11, 12), the intra-class correlation (configurations 13, 14), the correlation between random intercepts generating variables $x_1, x_2, x_3$ (configuration 15), the correlation between continuous variables in each cluster (configuration 16), the covariance matrix of the random effects (configuration 17, 18), the use of a probit link for generating binary covariates (configuration 19). Figure \ref{Figoverall} in Appendix \ref{Annexother} reports the distribution of the relative bias over the several configurations, while tables gathering inference results are available in the supplementary materials. We found similar results as compared to the base-case configuration.  
\section{Application to GREAT data}
\label{Application}
The MI methods are applied to the GREAT data (Appendix \ref{AnnexGREAT}). We considered model (\ref{analysiscont}) for analysis model, with $x_1$ representing the variable BNP, $x_2$ the variable AFIB and $y$ the variable LVEF. Although only three variables are included in the analysis model, the imputation models are based on nine variables to render the MAR assumption more credible \citep{Enders10,Schafer97}.

\begin{table}[H]
\centering
\begingroup\footnotesize
\caption{GREAT data: Point estimate, model standard error for parameters of a linear mixed effects model for several methods (Complete-case analysis, JM-jomo, FCS-GLM, FCS-2stage with REML estimator, FCS-2stage with moment estimator). 20 imputed arrays are considered for MI methods. 10 iterations are used for FCS methods. Time to multiply impute the dataset is indicated in minutes. Criteria related to the continuous (resp. binary) covariate are in light (resp. dark) grey.\label{resreal}} 
\begin{tabular}{rrrrrrr}
  \hline
 && \begin{sideways} CC \end{sideways} & \begin{sideways} JM-jomo \end{sideways} & \begin{sideways} FCS-GLM \end{sideways} & \begin{sideways} FCS-2stageREML \end{sideways} & \begin{sideways} FCS-2stageMM \end{sideways} \\ 
  \hline
\rowcolor[gray]{.9}&est & -0.1132 & -0.0891 & -0.1002 & -0.0854 & -0.1009 \\ 
 \rowcolor[gray]{.9}  \multirow{-2}{*}{$\beta_{BNP}$} &model se & 0.0108 & 0.0078 & 0.0163 & 0.0099 & 0.0112 \\ \hline 
\rowcolor[gray]{.7}  &est & 0.0268 & 0.0216 & 0.0218 & 0.0215 & 0.0273 \\ 
\rowcolor[gray]{.7}\multirow{-2}{*}{$\beta_{AFIB}$} &model se & 0.0071 & 0.0046 &0.0066 & 0.0040 & 0.0045 \\  \hline
 $\psi_{00}$&  est&            0.1112&  0.1075& 0.1232    &         0.1220&       0.1189\\ 
\rowcolor[gray]{.9}  $\psi_{BNP}$ &est& 0.0290 & 0.0306 & 0.0348& 0.0351 & 0.0332 \\ \hline
  time (min)&&  & 94.0 &30819.5 & 361.3 & 31.8 \\ 
   \hline
\end{tabular}
\endgroup
\end{table}

Results in Table \ref{resreal} indicate that the missing data mechanism in the GREAT application is not likely to be missing completely at random. In particular, complete-case analysis yielded estimates for the fixed coefficient $\beta_{BNP}$ farther away from null as compared to the MI methods.

As expected, standard errors obtained by CC were larger than those obtained from the base-case configuration of the simulation study. In general, standard errors for fixed effects estimates were smaller for the MI methods as compared to CC. An exception occurred for the variable $\beta_{BNP}$ when using FCS-2stageMM or FCS-GLM. Possibly, this is related to convergence issues resulting from limited number of iterations and relatively small number of imputed data sets. For instance, when we allowed for 50 iterations (rather than 10) and generated 50 imputed data sets (instead of 20), FCS-2stageMM yielded a standard error of 0.0091 for $\beta_{BNP}$.

Remarkably, we found that JM-jomo yielded the smallest standard errors for fixed effect estimates. This situation did not arise in the simulation studies, and is likely related to over-parametrisation of the FCS methods. In particular, the conditional imputation models assume random effects for all covariates, which leads to a substantial increase of the number of parameters as the number of covariates increases, hence inflating the variance around imputed values.

Finally, in agreement with the simulation studies, we found that the FCS-GLM method required substantially more computation time as compared to the other MI methods. This limitation is somewhat problematic, as the number of incomplete variables was rather limited in the GREAT application. As a result, checking convergence of the distribution of missing values to their posterior distribution becomes very difficult.

\section{Discussion}
\label{Conclusion}
As international collaboration becomes more common and access to large shared datasets increases, researchers increasingly often face incomplete multilevel data. Thus, handling systematically missing values becomes inevitable \citep{debray_individual_2015, Debray15}. In this work, we compared three recent multiple imputation methods for addressing this issue: JM-jomo, FCS-GLM and FCS-2stage. We also considered several extensions to better handle continuous and binary data in the presence of sporadically and systematically missing values. We highlighted the relevance of using these methods, and demonstrated their superiority over ad-hoc strategies through extensive simulation studies. Although the differences between the three imputation models are mainly technical, their properties may substantially differ according to the considered dataset.\\

In general, we found that JM-jomo tends to be conservative. This behaviour is in line with simulation study presented in \cite{Quartagno15}, and is related to the use of inverse-Wishart prior distributions for modelling the covariance matrices. Although this distribution avoids convergence issues of the Gibbs sampler \citep{Yucel02}, its use is not necessarily supported by the data. Furthermore, the prior distributions for the parameters of the inverse Wishart distribution appear to be rather influential. In particular, by sampling the covariance matrices using few degrees of freedom, too much variability is introduced for the within-cluster covariance matrices. As a result, fixed effects estimates vary too much across imputed datasets, leading to over-estimation of the variance components. Note that the influence of the prior distributions of imputation model's parameters have also been recently exhibited in the context of continuous data \cite{Kunkel17}. 

Bias is observed for JM-jomo when the number of individuals and/or clusters is small and/or variance of random effects is small. In such situations, the Inverse-Wishart prior distributions become very informative \citep{Gelman06}. Furthermore, because the Inverse-Wishart distribution tends to generate too much variability, its use may lead to shrinkage of regression coefficients when imputing continuous covariates ($\beta^{(1)}$ in the simulation study). This issue is less problematic for binary covariates ($\beta^{(2)}$)  because the diagonal terms of the within covariance matrices are constraint to be equal to one. However, inference for GLMM models with few clusters is challenging, even without missing data \citep{McNeish16cont}. When the number of clusters is large, substantial improvements can be obtained by increasing the degrees of freedom of the chi-squared distribution.\\

For FCS-GLM, we found that imputations were quite accurate for continuous variables. However, FCS-GLM is limited by the homoscedasticity assumption \citep{vanBuuren10,Resche16}. In particular, by fitting a homoscedastic model on heteroscedastic data, standard errors tend to be underestimated, even in the absence of missing data. As a result, the FCS-GLM method cannot fully propagate the sampling variability, leading to an underestimation of the variance of the parameters of the analysis model. This results in confidence intervals that are too narrow. However, the homoscedastic assumption can become a main advantage with small clusters since it avoids overfitting issues. As a result, the standard errors become well estimated also for continuous covariates. For this reason, FCS-GLM is a relevant method to use with small clusters. Another current problem of FCS-GLM is the time required for generating imputed datasets, particularly in large datasets. These results are in line with the simulation study of \cite{Resche16} comparing FCS-GLM and FCS-2stage.\\

FCS-2stage does not present any recurrent trend. Simulation study results suggest that using the log-normal distribution as approximation for the posterior distribution of the error variance outperformed modelling through Inverse-Wishart distributions (as in JM-jomo and FCS-GLM). Although the MM estimator of FCS-2stage is known to underestimate relevant variance components, similar inferences were obtained when adopting REML \citep{Langan16}. However, FCS-2stage may be problematic when imputing binary covariates in small clusters, as the maximum likelihood estimator used in stage one is known to yield biased estimates in such circumstances. For this reason, we investigated the use of Firth's correction \citep{Firth93}, but its implementation did not yield substantial improvements. Further research is warranted to investigate how FCS-2stage may be improved when applied to datasets with small clusters. In the GREAT application, we found that JM-jomo and FCS-2stage produce similar point estimates, but that the former yields smaller standard errors. This may reflect the ability of JM-jomo to borrow information about the study-specific covariance matrix across studies; and/or the appropriateness of using the Inverse Wishart model in the GREAT data.\\

A key issue in all MI methods is the use of \textit{congenial} imputation models \citep{Meng94}. Congeniality means that there is a joint model which implies the imputation model and the analysis model as submodels. Some results have been obtained for continuous variables when the analysis model does not include a random slope \citep{Quartagno15,Resche16}. However, as raised in \cite{Grund16}, with a random slope, these imputation models are uncongenial. Indeed, considering model (\ref{glm}), the outcome depends on a product of two random variables: the random effect ($b_k$) and the associated covariate ($\bfW_k$). Consequently, the marginal distribution of the outcome becomes highly complex, whereas a joint imputation model like the one used in JM-jomo (without covariate in the right part of model (\ref{glmmulti})), assumes simpler Gaussian marginal distributions in each cluster. In the same way, for FCS methods, the conditional distribution of one covariate is no longer analytically tractable. This implies that the distribution of the covariates given the outcome cannot be written according to a GLMM model. Thus, imputation models are misspecified whatever the imputation method used. Nevertheless, our simulation study highlights that this is a minor practical issue since the biases remain very small for fixed coefficients and variance of random effects (see also Appendix \ref{newappendixcong}). Recent progress to provide imputation models ensuring congeniality even in complex settings \citep{Bartlett15} seems promising for the multilevel setting.

Another source of mis-specification is the choice of random and fixed effects in the imputation models. In particular, tuning each conditional imputation model is tricky in practice for FCS approaches, particularly with a lot of variables, although models can become over-parametrised otherwise. More generally, finding conditional imputation models with few parameters in FCS approaches is a current topic of research \citep{Zhao16} in the one-level case, and appears even more challenging in the two-level approach. In this paper we used the default model of the methods: for JM-jomo, all variables are in the response part of model (2), which corresponds to normal marginal distributions with random intercept, whereas FCS approaches include all covariates in fixed and random effects, making such marginal distribution more complex. Consequently, the imputation models are not strictly the same according to this aspect, that could also explain some differences between JM and FCS approaches.

An additional difficulty for all MI methods is the imputation of binary variables. Whereas JM-jomo overcomes it quite well by considering a fully Bayesian multilevel modelling approach with a probit link function, FCS-GLM and FCS-2stage are less tailored for such variables: FCS-GLM because it considers a logit link making difficult to handle sporadically missing values, and FCS-2stage because it infers separately on each clusters, making samples too small to provide accurate inferences. For these reasons, both FCS methods could be improved by adopting: a probit link function for FCS-GLM, and applying bias correction for variances and point estimates for FCS-2stage. Note that in contrast to FCS-2stage, the methods FCS-GLM and JM-jomo can also handle nominal and count variables. FCS-2stage could further be extended by considering additional link functions in the regression models used at stage 1. Finally, although FCS-2lnorm provides encouraging performance for imputing missing continuous and binary multilevel data, it does not \textit{properly} reflect \citep[p.~105]{Schafer97} the variability of random coefficients. For this reason, its usefulness remains limited in the presence of systematically missing values. 

Although we only considered a 2-level setting in this study, extensions of the presented MI methods to a higher hierarchical structure are relatively straightforward. At this moment, only JM-jomo proposes an implemented solution to address such situations. Note that missing values may also occur at level-2. JM-jomo naturally handles this setting, but FCS approaches have also been developed to achieve this goal \citep{mice}. In addition, we did not focus the topic on longitudinal data, or more generally on data with very few observations per cluster, like in education research field. However, systematically missing values are also frequent in such cases and raise additional overfitting issues of the imputation models.

Furthermore, our study focuses on the use of GLMM models to analyse multilevel data, which facilitates the use of MI. Indeed, direct maximum likelihood inference is another possible strategy to address missing data, but its implementation becomes difficult when dealing with multilevel variables \citep{Longford08, Schafer97}. However, many other statistics than the parameters of a GLMM model can be of interest. For instance, \cite{Curran09,Curran08} proposed using item response theory to fit measurement models.\\ 

From a general point of view, whatever the imputation method used, accurate inferences for a GLMM model can be expected only with a high (or moderate) number of clusters. Heteroscedastic MI methods perform better than homoscedastic methods, which should be reserved with few individuals only. Methods based on conjugate prior distributions should be used with caution when the proportion of missing values is very high. Multiple imputation of binary variable is challenging, and methods can have drawbacks in such a case.

In this regards, JM-jomo could be recommended when the number of incomplete binary variables is large and when the number of observed clusters is large. FCS-2stage performs quite well, but should be avoided when clusters are small, or equivalently, when the proportion of sporadically missing values is large. This method is particularly relevant compared to the others when the number clusters with systematically missing variables is large. The MM version offers a quick solution to have an initial overview of the inference results. Finally, FCS-GLM appears advantageous when clusters are small.

From our point of view, the topic of inference for multilevel incomplete data needs to strengthen the theoretical underpinnings to improve the fit of imputation models, as well as some developments to broaden the scope of the evaluated methods. Among them, the congeniality has been recently discussed but need more efforts for analysis model with random slope. Machine-learning methods offering more flexibility could be considered for this purpose. In addition, solutions to handle missing data without assuming the ignorability of the missing data mechanism need to be investigated. Furthermore, in the big data era, MI methods handling a large number of variables need also to be studied. Finally, proposing imputation models for nominal or ordered variables avoiding informative prior distributions constitutes a main line of research.

\section*{Acknowledgements}
We thank the Global Research on Acute Conditions Team \citep{great} for allowing the use of their data.
\section*{Funding}
TPAD was supported by the Netherlands Organization for Scientific Research (91617050 and 91215058).
IRW was funded by the Medical Research Council [Unit Programme number U105260558 and MC\_ UU\_ 12023/21].
VA and MRR was supported by the Agence nationale de s\'ecurit\'e du m\'edicament et des produits de sant\'e (ANSM).
\bibliographystyle{apalike}
\bibliography{biblio}
\appendix

\section{Posterior distributions}
\label{pd_annexe}
This section reports the prior and posterior distributions of the parameters of MI method as well as technical points to obtain realisations from them.
\subsection{JM-jomo\label{pd_jomo}}
The prior distributions for parameters $\PI=\left(\bfbeta,\bfPsi,\left(\bfSigma_{\nbclust}\right)_{1\leq\nbclust\leq\Nbclust}\right)$ of model (\ref{glmmulti}) are as follows:
\begin{eqnarray*}
\bfbeta\propto 1\\
\bfPsi^{-1}\sim \mathcal{W}\left(\nu_1,\bfLambda_1\right)\\
\bfSigma_{\nbclust}^{-1}\vert\nu_2,\bfLambda_2 \sim \mathcal{W}\left(\nu_2,\bfLambda_2\right) \text{for all } {1\leq\nbclust\leq\Nbclust} \\
\nu_2 \sim \chi^2\left(\eta\right),\ \bfLambda_2^{-1} \sim \mathcal{W}\left(\nu_3,\bfLambda_3\right)\label{eqnjm}
\end{eqnarray*}
where $\mathcal{W}\left(\nu,\bfLambda\right)$ denotes the Wishart distribution with $\nu$ degrees of freedom and scale matrix  $\bfLambda$, and $\eta$ denotes the degrees of freedom of the chi-squared distribution. Hyperparameters are set as $\nu_1=\Nbout$, $\bfLambda_1=\mathbb{I}_{\Nbout}$, $\nu_3=\Nbout\Nbclust$, $\bfLambda_3=\mathbb{I}_{\Nbout\Nbclust}$ and $\eta=\Nbout\Nbclust$. 

If data were complete, conditional posterior distributions for each component of $\PI$ would be:
\begin{eqnarray}
\bfbeta\vert Y, \left(\bfSigma_\nbclust\right)_{1\leq\nbclust\leq\Nbclust},b \sim \mathcal{N}\left(\widehat{\bfbeta},\widehat{\text{var}}\left(\widehat{\bfbeta}\right)\right) \label{jm1}\\
\bfPsi^{-1}\vert Y,b \sim \mathcal{W}\left(\nu_1+\Nbclust,\left(\bfLambda_1^{-1}+S_1\right)^{-1}\right) \\ \label{jm2}
\bfSigma_{\nbclust}^{-1}\vert Y,b,\bfbeta,\nu_2,\bfLambda_2 \sim \mathcal{W}\left(\nu_2+\Nbind,\left(\bfLambda_2^{-1}+S_2\right)^{-1}\right)\label{jm3}
\end{eqnarray}
with $\widehat{\bfbeta}$ the estimate of the weighted least-squares estimator, $\widehat{\text{var}}\left(\widehat{\bfbeta}\right)$ the estimate of the associated variance, $S_1=\left(b_1^V,\ldots, b_\Nbclust^{V}\right)\left(b_1^V,\ldots, b_\Nbclust^{V}\right)^{\top}$, $S_2=\sum_\nbclust {\widehat{\varepsilon}}^{V}_\nbclust\left( {\widehat{\varepsilon}_\nbclust}^{V}\right)^{\top}$ where $\widehat{\varepsilon}_{\nbclust}$ denotes the residuals for cluster $\nbclust$.

To draw the parameters of the imputation model from their posterior distribution with a multivariate missing data pattern, a DA algorithm is used \citep{Tanner87}. Note that the posterior distribution of $\PI$ also depends on the random coefficients and on the random parameters $\left(\nu_2, \bfLambda_2\right)$, which is why they are updated at each cycle of the algorithm.\\

To deal with binary variables, a probit link and a latent variables framework have been proposed \citep{Goldstein09}. Note that drawing $\left(\bfSigma_\nbclust\right)_{1\leq\nbclust\leq\Nbclust}$ from its posterior distribution is more tricky with binary variables since $\bfSigma_{\nbclust}$ has to respect the constraint of diagonal elements equal to one for latent variables \citep{Browne06}.\\

\subsection{FCS-GLM\label{pd_glm}}
For continuous incomplete variables, the conditional imputation model is model (\ref{glm}) assuming homoscedastic error terms, \textit{i.e.} $\sigma_\nbclust=\sigma$ for all $\nbclust$. From non-informative independent priors, components of the parameter $\PI=\left(\bfbeta, \bfPsi, \sigma\right)$ are drawn from the following posterior distributions: 
\begin{eqnarray}
\sigma_{\nbcol}^{2}\vert Y, b \sim \text{Inv-}\Gamma\left(\frac{\Nbind-\Nbvar}{2},\frac{\left(\Nbind-\Nbvar\right)\widehat{\sigma^2}}{2}\right)\label{fcs3}\\
\bfbeta\vert Y,b,\sigma_{\nbcol}^{2} \sim \mathcal{N}\left(\widehat{\bfbeta},\widehat{\text{var}}\left(\widehat{\bfbeta}\right)\right)\label{fcs1} \\
\bfPsi^{-1}\vert Y, b \sim \mathcal{W}\left(\Nbclust,\widehat{b}\widehat{b}^{\top}\right) \label{fcs2}
\end{eqnarray}
where $\text{Inv-}\Gamma\left(s,r\right)$ denotes the inverse Gamma distribution with shape $s$ and rate $r$, $\widehat{\bfbeta}$ and $\widehat{\sigma^2}$ are the ML estimates of $\bfbeta$ and $\sigma^2$, $\widehat{\text{var}}\left(\widehat{\bfbeta}\right)$ is the variance estimate associated with the ML estimator of $\bfbeta$, and $\widehat{b}=\left(\widehat{b}_1,\dots, \widehat{b}_{\Nbclust}\right)$ are the best linear unbiased predictions (BLUP) of the random effects, preferably estimated by using the restricted ML (REML) estimator \cite{Jolani15}.

 With systematically missing values on variable $y$, ML estimates are evaluated from the clusters observed in $y$ and degrees of freedom in (\ref{fcs2}) and (\ref{fcs3}) are modified accordingly.\\

Note that posterior distributions are conditional to the random coefficients that are unknown. Rigorously, simulation of these posterior distributions require the use of a Gibbs sampler. To avoid the use of such iterative algorithm, the marginal distribution of $\PI$ is approximated by the one of $\PI$ given a fixed value for $b$ (i.e. $P\left(\PI\vert Y\right)\approx P\left(\PI\vert Y;b=\hat{b}\right)$). Assuming having mimicked the marginal posterior distribution of $\theta$ by making one draw, then we can draw random effects given this one. Furthermore, contrary to the original paper \citep{Jolani15} the prior distribution accounts for the dependence between the components $\sigma^2$ and $\beta$ \citep{Jolani17}. \\

Binary variables are imputed in the same way by considering a GLMM model with a logit link.\\

With sporadically missing continuous variables, steps (\ref{fcs3}-\ref{fcs2}) are essentially the same: the parameters of the posterior distribution are tuned by considering the ML estimator evaluated from individuals observed on variable $\nbvar$ and degrees of freedom are suitably modified in (\ref{fcs3}).
\subsection{FCS-2step\label{pd_2step}}
For a continuous incomplete variable $y$, FCS-2stage is based on the conditional imputation model:

\begin{eqnarray*}
y_{\nbclust}&=&\bfZ_{\nbclust}\left(\bfbeta+b_{\nbclust}\right)+\varepsilon_{\nbclust}\\
b_{\nbclust}&\sim& \mathcal{N}\left(0,\bfPsi\right) \nonumber\\
\varepsilon_{\nbclust} &\sim& \mathcal{N}\left(0,\sigma^2_{\nbclust}\mathbb{I}_{\Nbind_\nbclust}\right)\nonumber
\end{eqnarray*}
The parameter of this model is $\PI=\left(\bfbeta,\bfPsi,\left(\sigma_{\nbclust}\right)_{1\leq\nbclust\leq\Nbclust}\right)$. 

FCS-2stage draws the parameters of the imputation model by using an asymptotic strategy: a two-stage estimator is evaluated from the observed data and the posterior distribution is then approximated.
To fit the two-stage estimator, at stage one, the ML estimator of a linear model is computed on each available cluster:
\begin{eqnarray*}
\widehat{\bfbeta}_{\nbclust}=\left(\bfZ_{\nbclust}^{\top}\bfZ_{\nbclust}\right)^{-1}\bfZ_{\nbclust}^{\top}\bfy_{\nbclust}
\end{eqnarray*}
Then, at stage two, the following random effects model is used:
\begin{eqnarray*}
\widehat{\bfbeta}_\nbclust=\bfbeta+b_{\nbclust}+\varepsilon'_{\nbclust}
\end{eqnarray*}
with $b_{\nbclust}\sim \mathcal{N}\left(0,\bfPsi\right)$ and $\varepsilon'_{\nbclust}\sim\mathcal{N}\left(0,\sigma_\nbclust^2\left(\bfZ_{\nbclust}\bfZ_{\nbclust}^{\top}\right)^{-1}\right)$.
$\bfbeta$ and $\bfPsi$ may be estimated by REML; alternatively, \cite{Resche16} suggest using the method of moments (MM), which is even faster, especially with high dimensional $\bfbeta$ \citep{derSimonian86,Jackson13}.

The two-stage estimator is typically used by assuming $\left(\sigma_{\nbclust}\right)_{1\leq\nbclust\leq\Nbclust}$ to be known \citep{Burke16}. In our framework, $\left(\sigma_{\nbclust}\right)_{1\leq\nbclust\leq\Nbclust}$ is unknown and cannot be identified at stage one with systematically missing values. Thus, instead of fixing $\sigma_{\nbclust}$ $\left(1\leq\nbclust\leq\Nbclust\right)$, the assumption is made that $\left(\sigma_{\nbclust}\right)_{1\leq\nbclust\leq\Nbclust}$ are the independent realisations of a unique random variable so that
\begin{eqnarray}\label{distrisigma}
\text{log} \ \sigma_{\nbclust}\sim \mathcal{N}\left(\text{log}\ \sigma,\Phi\right).
\end{eqnarray}
As with estimation of $\bfbeta$ and $\bfPsi$ in (\ref{2stage-1}-\ref{2stage}), $\text{log}\ \sigma$ and $\Phi$ are estimated by a two-stage method. At stage one, the ML estimate of $\sigma_{\nbclust}$, denoted  $\widehat{\sigma}_{\nbclust}$, is computed and log-transformed on each cluster, and its variance is derived (using the Delta method). At stage two, these estimates are used in the random effects model
\begin{eqnarray}\label{2stagesigma}
\text{log}\ \widehat{\sigma}_{\nbclust}=\text{log}\ \sigma +s_{\nbclust}+\varepsilon''_{\nbclust}
\end{eqnarray}
with $s_\nbclust\sim \mathcal{N}\left(0,\Phi\right)$, $\varepsilon''_{\nbclust}\sim \mathcal{N}\left(0,\text{var}\left(\text{log}\ \widehat{\sigma}_{\nbclust}\right)\right)$.

To draw missing values of $y$ from their predictive distribution using this estimator, first, $\text{log}\ \sigma$ and $\Phi$ are estimated by fitting the two-stage estimator to the data and their posterior distribution is then approximated by
\begin{eqnarray}
\text{log}\ \sigma\vert Y^{\text{obs}} \sim \mathcal{N}\left(\widehat{\text{log}\ \sigma},\widehat{\text{var}}\left(\widehat{\text{log}\ \sigma}\right)\right)\label{fcs2step1}\\
\Phi\vert Y^{\text{obs}} \sim \mathcal{N}\left(\widehat{\Phi},\widehat{\text{var}}\left(\widehat{\Phi}\right)\right).
\end{eqnarray}
Next, from model (\ref{2stagesigma}), $s_\nbclust$ can be straightforwardly drawn conditionally on $\text{log}\ \widehat{\sigma}_{\nbclust}$ since $\text{log}\ \widehat{\sigma}_{\nbclust}$ and $s_{\nbclust}$ are Gaussian, providing values for $\left(\sigma_{\nbclust}\right)_{1\leq\nbclust\leq\Nbclust}$. Then, $\bfbeta$ and $\bfPsi$ are drawn from their posterior distribution in the same way:
\begin{eqnarray}
\bfbeta\vert Y^{\text{obs}}, \sigma_{\nbclust} \sim \mathcal{N}\left(\widehat{\bfbeta},\sigma_\nbclust^2\left(\bfZ_{\nbclust}\bfZ_{\nbclust}^{\top}\right)^{-1}\right)\label{fcs2step3}\\
\bfPsi^{\text{chol}}\vert Y^{\text{obs}}, \sigma_{\nbclust} \sim \mathcal{N}\left(\widehat{\bfPsi}^{\text{chol}},\widehat{\text{var}}\left(\widehat{\bfPsi}^{\text{chol}}\right)\right)\label{fcs2step4}
\end{eqnarray}
where $\bfPsi^{\text{chol}}$ is the Cholesky decomposition of $\bfPsi$.

To handle binary variables with both sporadically and systematically missing values by applying a logit link in the imputation model (\ref{model2step}). In this case the parameters $\sigma_k$ are not used, thus the distribution assumption (\ref{distrisigma}) does not hold and the two-stage estimator described in step (\ref{2stagesigma}) is not needed.

\section{Description of GREAT data}\label{AnnexGREAT}
The GREAT Network performed an IPD meta-analysis to explore risk factors associated with short-term mortality in acute heart failure (AHF) \citep{great}. Their dataset consists of 28 studies: 8 were carried out in Western Europe (2 in Italy, 2 in Spain, and 1 in each of France, Finland, Switzerland, Netherlands), 13 in Central Europe (12 in Czech Republic and 1 in Austria), 3 in America (2 in the United States and 1 in Argentina), 3 in Asia (China, Japan, Korea), and 1 in Africa (Tunisia) \citep{Mebazaa13}. The principal investigators of each study provided the original data collected for each patient, including a list of patient characteristics and potential risk factors \citep{Lassus13}.

One biomarker of interest was brain natriuretic peptide (BNP), which is known to be elevated in acute heart failure. Since measuring the left ventricular ejection fraction (LVEF) requires an ultrasound examination, one objective consists in explaining LVEF from biomarkers that are easier to measure, such as BNP or electrocardiographic characteristics such as the atrial fibrillation (AFIB). The generalized linear mixed effect model (GLMM)\citep{Pinheiro00, Lee06} is a suitable statistical model to achieve this goal.

The dataset contains 2 binary variables (AFIB and Gender) and 7 continuous variables (BMI, Age, Systolic blood pressure (SBP), Diastolic blood pressure (DBP), Heart rate (HR), LVEF and BNP). Variables are described in Table \ref{tabledon} in Appendix \ref{AnnexGREAT}. The total number of individuals is 11685 and study sizes range from 18 to 1834.

Each study is incomplete, leading to sporadically missing values on all variables except gender and LVEF. However, BNP measurement is a recent technique, so this variable has been collected on 10 studies only, leading to systematically missing values. Four other variables are systematically missing on some studies (Table \ref{tablena}), notably the binary variable AFIB.
\begin{table}[H]
\begin{center}
\caption{\label{tablena} GREAT data: percentages of missing values by variable and study.}
\scriptsize{\begin{tabular}{rrrrrrrrrrr}
  \hline
 $\nbclust$ & $\Nbind_{\nbclust}$ & Gender & BMI & Age & SBP & DBP & HR & BNP & AFIB & LVEF \\ 
   \hline
1 & 410 & 0 & 36 & $<$ 1 & 1 & 2 & 3 & 57 & $<$ 1 & 0 \\ 
  2 & 567 & 0 & 19 & 0 & 2 & 3 & 1 & 10 & 0 & 0 \\ 
  3 & 210 & 0 & 43 & 0 & 1 & 2 & 1 & 0 & 100 & 0 \\ 
  4 & 375 & 0 & 2 & 0 & 1 & 1 & 2 & 4 & 42 & 0 \\ 
  5 & 107 & 0 & 1 & 0 & 0 & 0 & 0 & 100 & 0 & 0 \\ 
  6 & 267 & 0 & 100 & 0 & 100 & 100 & $<$ 1 & 100 & 0 & 0 \\ 
  7 & 203 & 0 & $<$ 1 & 0 & 1 & 2 & 1 & $<$ 1 & 0 & 0 \\ 
  8 & 354 & 0 & 44 & 1 & 16 & 16 & 19 & 12 & 22 & 0 \\ 
  9 & 137 & 0 & 1 & 0 & 0 & 0 & 0 & 0 & 0 & 0 \\ 
  10 & 48 & 0 & 100 & 0 & 0 & 0 & 4 & 100 & 0 & 0 \\ 
  11 & 208 & 0 & 24 & 0 & 0 & $<$ 1 & 0 & 100 & 0 & 0 \\ 
  12 & 622 & 0 & 27 & 0 & $<$ 1 & $<$ 1 & 1 & 100 & 0 & 0 \\ 
  13 & 78 & 0 & 60 & 0 & 0 & 0 & 0 & 100 & 100 & 0 \\ 
  14 & 670 & 0 & 77 & $<$ 1 & 1 & 1 & 2 & 100 & $<$ 1 & 0 \\ 
  15 & 1000 & 0 & 13 & 0 & 2 & 2 & 2 & 82 & $<$ 1 & 0 \\ 
  16 & 1093 & 0 & 0 & 0 & 0 & 0 & 0 & 100 & 0 & 0 \\ 
  17 & 18 & 0 & 6 & 0 & 0 & 0 & 0 & 22 & 0 & 0 \\ 
  18 & 1834 & 0 & 19 & 0 & 1 & 1 & $<$ 1 & 92 & $<$ 1 & 0 \\ 
  19 & 358 & 0 & 7 & 0 & 0 & 0 & 0 & 99 & 0 & 0 \\ 
  20 & 54 & 0 & 6 & 0 & 2 & 2 & 2 & 100 & 2 & 0 \\ 
  21 & 588 & 0 & 10 & 0 & $<$ 1 & $<$ 1 & 0 & 97 & $<$ 1 & 0 \\ 
  22 & 651 & 0 & 24 & 0 & 2 & 2 & 2 & 73 & 2 & 0 \\ 
  23 & 455 & 0 & 2 & 0 & 0 & 0 & $<$ 1 & 86 & $<$ 1 & 0 \\ 
  24 & 294 & 0 & 4 & 0 & $<$ 1 & $<$ 1 & $<$ 1 & 81 & 0 & 0 \\ 
  25 & 397 & 0 & 1 & 0 & 0 & 0 & 0 & 100 & 0 & 0 \\ 
  26 & 295 & 0 & 11 & 0 & 0 & 0 & 0 & 66 & 0 & 0 \\ 
  27 & 303 & 0 & 11 & 0 & $<$ 1 & $<$ 1 & 0 & 79 & 0 & 0 \\ 
  28 & 89 & 0 & 0 & 0 & 0 & 0 & 0 & 38 & 0 & 0 \\ 
   \hline
\end{tabular}}
\end{center}
\end{table}

\begin{table}[H]
\begin{center}
\caption{\label{tabledon} GREAT data: description of variables. Binary variables are presented by counts and percentages, while continuous variables by their median and quartiles.}
\begin{tabular}{lrrl}
\hline
variable&value&size&summary\\
\hline gender&0&6865&58.65\%\\
&1&4820&42.35\%\\
AFIB&0&7704&69.18\%\\
&1&3431&30.81\%\\
BMI&& 9259 &26.58  [ 23.66143862 ; 30.12 ]\\
Age&& 11678  &72.7  [ 62.7 ; 80 ]\\
SBP&&11278  &130  [ 111 ; 153 ]\\
DBP&&11262 &80  [ 68 ; 90 ]\\
HR&& 11518 &87  [ 72 ; 105 ]\\
LVEF&& 11685 &0.38 [ 0.27 ; 0.5 ]\\
BNP&&  2776&2.99 [ 2.66 ; 3.29 ]\\ \hline
\end{tabular}
\end{center}
\end{table}

\section{Simulation}\label{AnnexSimu}

\subsection{Simulation design}
\subsubsection{Investigated configurations}\label{AnnexDesign}
~~
\begin{sidewaystable}
\begin{center}
\scriptsize{
\caption{Tuning parameters for the several configurations: $\Nbclust$ the number of clusters, $\Nbind$ the number of individuals, $\lambda$ a scalar multiplying the variance of random effects, $\rho_{v}$ the correlation between random intercepts generating variables $x_1, x_2, x_3$, $\rho_{b}$ the correlation between random coefficients of the analysis model, the type of the outcome, the type of the covariates ($x_1$ or $x_2$), the proportion of systematically missing values on the covariates ($x_1$ or $x_2$), the proportion of sporadically missing values on the covariates ($x_1$ or $x_2$), the nature of the missing data mechanism $R$, the intra cluster correlation for continuous variables (ICC), $\rho_{x_{1},x_{3}}$ the correlation between $x_1$ and $x_3$ in each cluster, the presence of a random effect on the covariate $x_2$, the link function used to generate the binary covariate. Parameters varying from the base-case configuration are in boldface.\label{summarysimu}}
\begin{tabular}{l|lllll|l|llllll|llllll}
 \hline \multirow{3}{*}{case} & \multirow{3}{*}{$\Nbclust$} &  \multirow{3}{*}{$\Nbind$}&  \multirow{3}{*}{$\lambda$}&  \multirow{3}{*}{{$\rho_{v}$}}&  \multirow{3}{*}{{$\rho_b$}}&\multicolumn{1}{c|}{$y$}&\multicolumn{6}{c|}{$x_1$}&\multicolumn{6}{c}{$x_2$}
 \\ \cline{7-19}
&  &&&  &&\rotatebox[]{90}{type}&\rotatebox[]{90}{type}&\rotatebox[]{90}{$\pi_{syst}$}&\rotatebox[]{90}{$\pi_{spor}$}&\rotatebox[]{90}{$R$}&\rotatebox[]{90}{ICC}&\rotatebox[]{90}{$\rho\left(x_{1},x_{3}\right)$}&\rotatebox[]{90}{type}&\rotatebox[]{90}{$\pi_{syst}$}&\rotatebox[]{90}{$\pi_{spor}$}&\rotatebox[]{90}{$R$}&\rotatebox[]{90}{random effect on $x_2$}&\rotatebox[]{90}{link function}\\ \hline
1 (base-case)&28& 11685&1&.07&-.87&cont&cont&.25&.25&MCAR&.25&.3&bin&.25&.25&MCAR&no&logistic\\
2&28& 11685&1&.07&-.87&cont&cont&\textbf{.1}&\textbf{.375}&MCAR&.25&.3&bin&\textbf{.1}&\textbf{.375}&MCAR&no&logistic\\
3&28& 11685&1&.07&-.87&cont&cont&\textbf{.4}&\textbf{.0625}&MCAR&.25&.3&bin&\textbf{.4}&\textbf{.0625}&MCAR&no&logistic \\
4&28& 11685&1&.07&-.87&cont&\textbf{bin}&.25&.25&MCAR&&&bin&.25&.25&MCAR&no&logistic \\
5&28& 11685&1&.07&-.87&\textbf{bin}&cont&.25&.25&MCAR&.25&.3&bin&.25&.25&MCAR&no&logistic\\ 
6&28& 11685&1&.07&-.87&cont&cont&.25&.25&MCAR&.25&.3&bin&\textbf{0}&\textbf{0}&&no&logistic\\
7&28& 11685&1&.07&-.87&cont&cont&\textbf{0}&\textbf{0}&&.25&.3&bin&.25&.25&MCAR&no&logistic\\
8&28& 11685&1&.07&-.87&cont&cont&.25&.25&\textbf{MAR}&.25&.3&bin&.25&.25&\textbf{MAR}&no&logistic\\
9&28& 11685&1&.07&-.87&cont&cont&.25&.25&MCAR&.25&.3&bin&.25&.25&MCAR&\textbf{yes}&logistic\\
10&\textbf{14}& 11685&1&.07&-.87&cont&cont&.25&.25&MCAR&.25&.3&bin&.25&.25&MCAR&no&logistic\\
11&28& \textbf{5845}&1&.07&-.87&cont&cont&.25&.25&MCAR&.25&.3&bin&.25&.25&MCAR&no&logistic\\
12&28& \textbf{2923}&1&.07&-.87&cont&cont&.25&.25&MCAR&.25&.3&bin&.25&.25&MCAR&no&logistic\\
13&28& 11685&1&.07&-.87&cont&cont&.25&.25&MCAR&\textbf{.5}&.3&bin&.25&.25&MCAR&no&logistic\\
14&28& 11685&1&.07&-.87&cont&cont&.25&.25&MCAR&\textbf{.1}&.3&bin&.25&.25&MCAR&no&logistic\\
15&28& 11685&1&\textbf{.3}&-.87&cont&cont&.25&.25&MCAR&.25&.3&bin&.25&.25&MCAR&no&logistic\\
16&28& 11685&1&.07&-.87&cont&cont&.25&.25&MCAR&.25&\textbf{.5}&bin&.25&.25&MCAR&no&logistic\\
17&28& 11685&\textbf{2}&.07&-.87&cont&cont&.25&.25&MCAR&.25&.3&bin&.25&.25&MCAR&no&logistic\\
18&28& 11685&1&.07&\textbf{-.3}&cont&cont&.25&.25&MCAR&.25&.3&bin&.25&.25&MCAR&no&logistic\\
19&28& 11685&1&.07&-.87&cont&cont&.25&.25&MCAR&.25&.3&bin&.25&.25&MCAR&no&\textbf{probit}\\
20&\textbf{7}& 11685&1&.07&-.87&cont&cont&.25&.25&MCAR&.25&.3&bin&.25&.25&MCAR&no&logistic\\
\hline
\end{tabular}}
\end{center}
\end{sidewaystable}
\newpage

\subsection{Complementary results}\label{AnnexResults}

\subsubsection{Robustness to the proportion of systematically missing values}\label{AnnexSyst}
~~
\begin{figure}[H]
\begin{center}
\caption{Robustness to the proportion of systematically missing values: relative bias for the estimate of $\beta^{(1)}$ (left), $\beta^{(2)}$ (middle) and $\psi_{11}$ (right) according to $\pi_{syst}$ for each MI method. The proportion of sporadically missing values is modified to keep a constant proportion of missing values (in expectation).\label{figa1}}
\includegraphics[width=4cm, height=6cm]{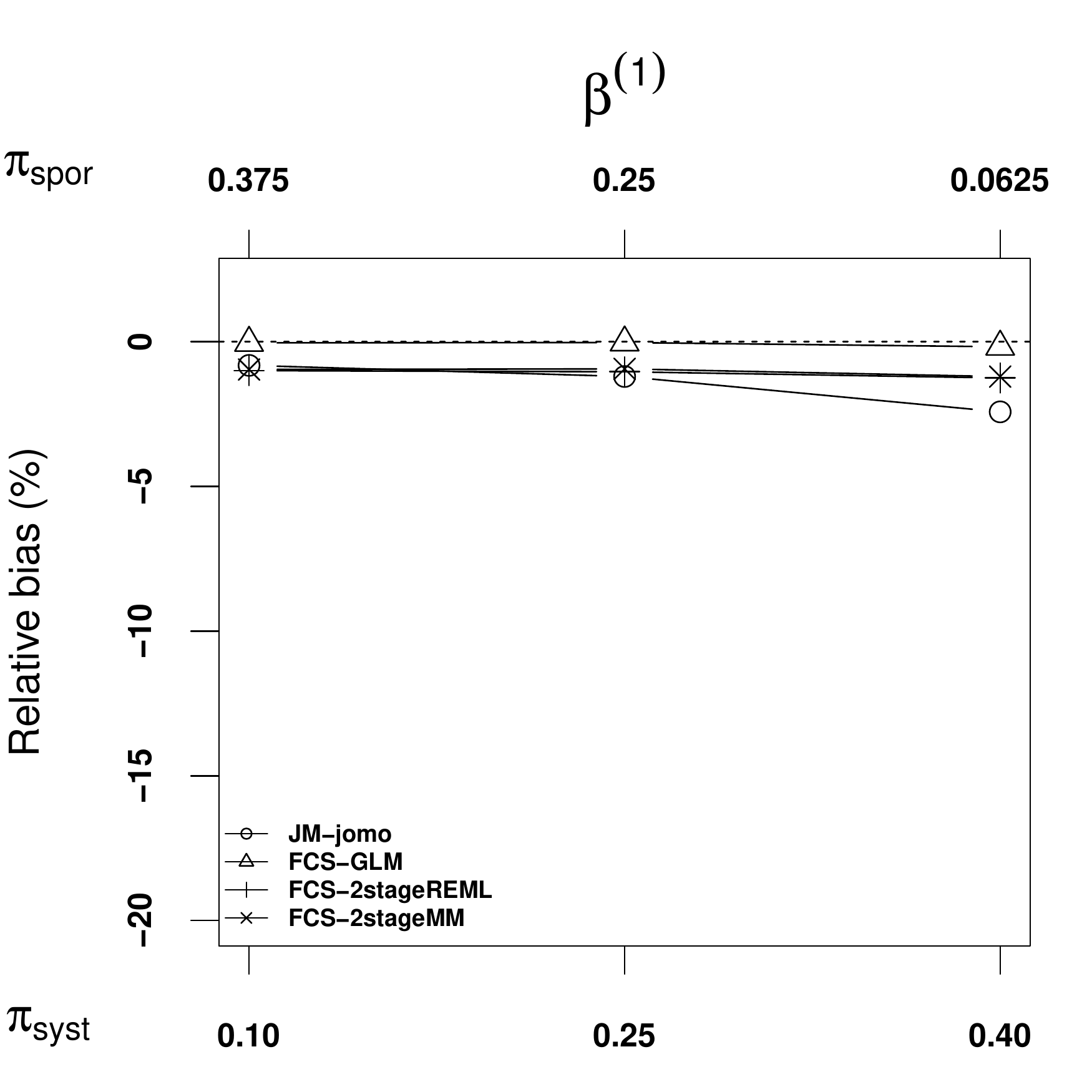} \includegraphics[width=4cm, height=6cm]{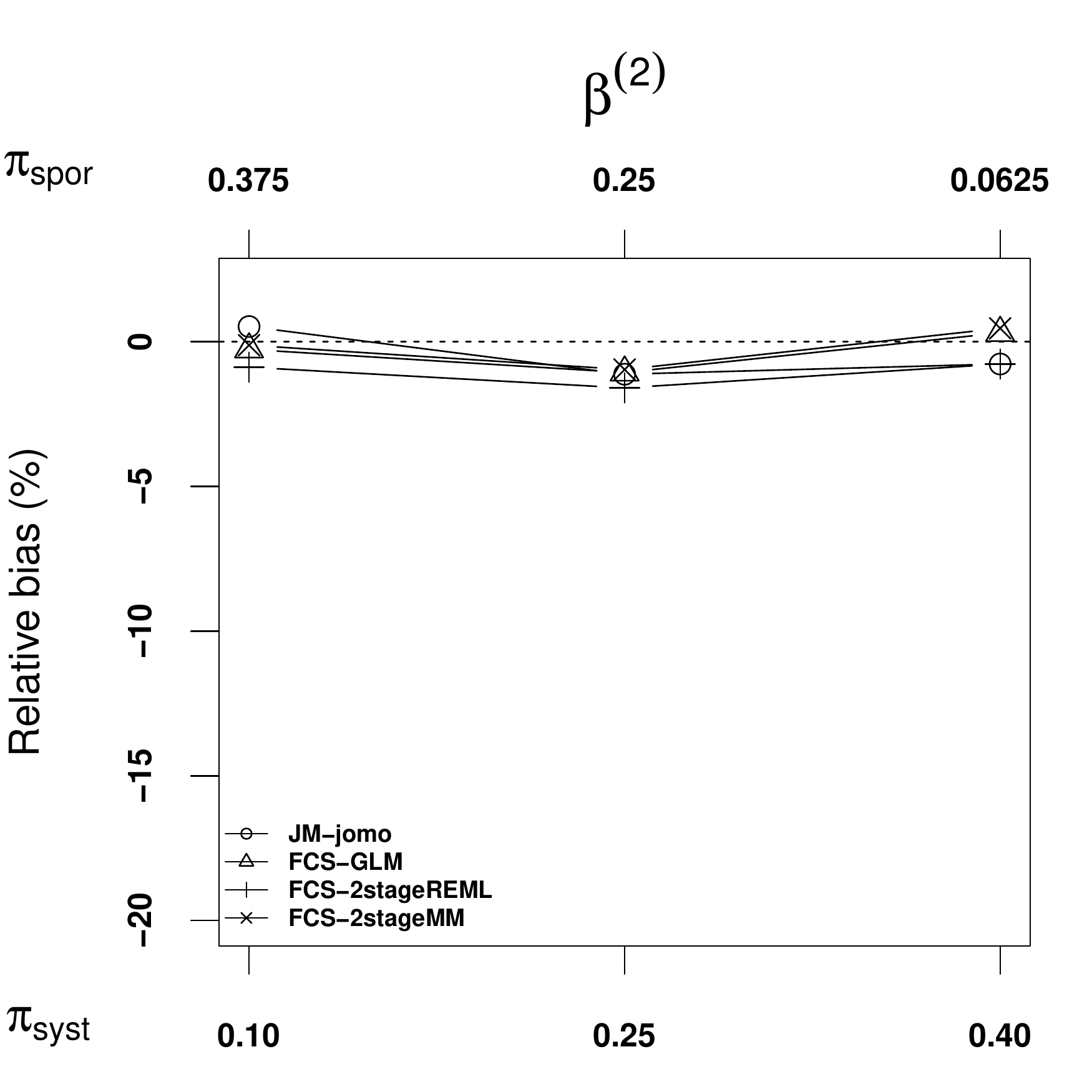}\includegraphics[width=4cm, height=6cm]{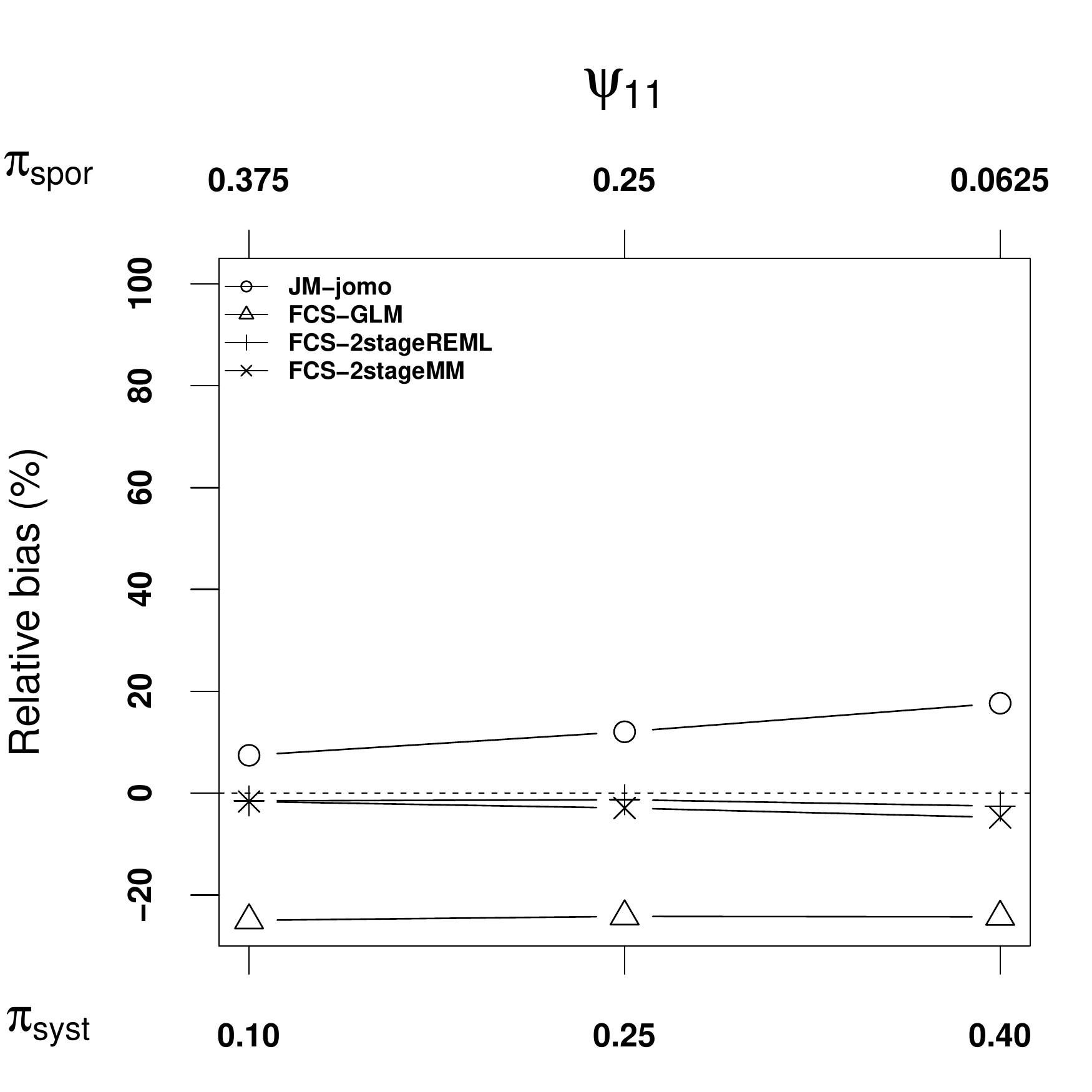}
\end{center}
\end{figure}

\subsubsection{Robustness to the number of clusters}\label{AnnexK}
~~
\begin{figure}[H]
\begin{center}
\caption{Robustness to the number of clusters: estimate of the relative bias for the SE estimate for $\widehat{\beta^{(1)}}$ (left), $\widehat{\beta^{(2)}}$ (right) according to $\Nbclust$ for each MI method.  The estimated relative bias is calculated by the difference between the model SE and the empirical SE, divided by the empirical SE. Criteria are based on 500 incomplete datasets.\label{figa2}}
\includegraphics[width=6cm, height=8cm]{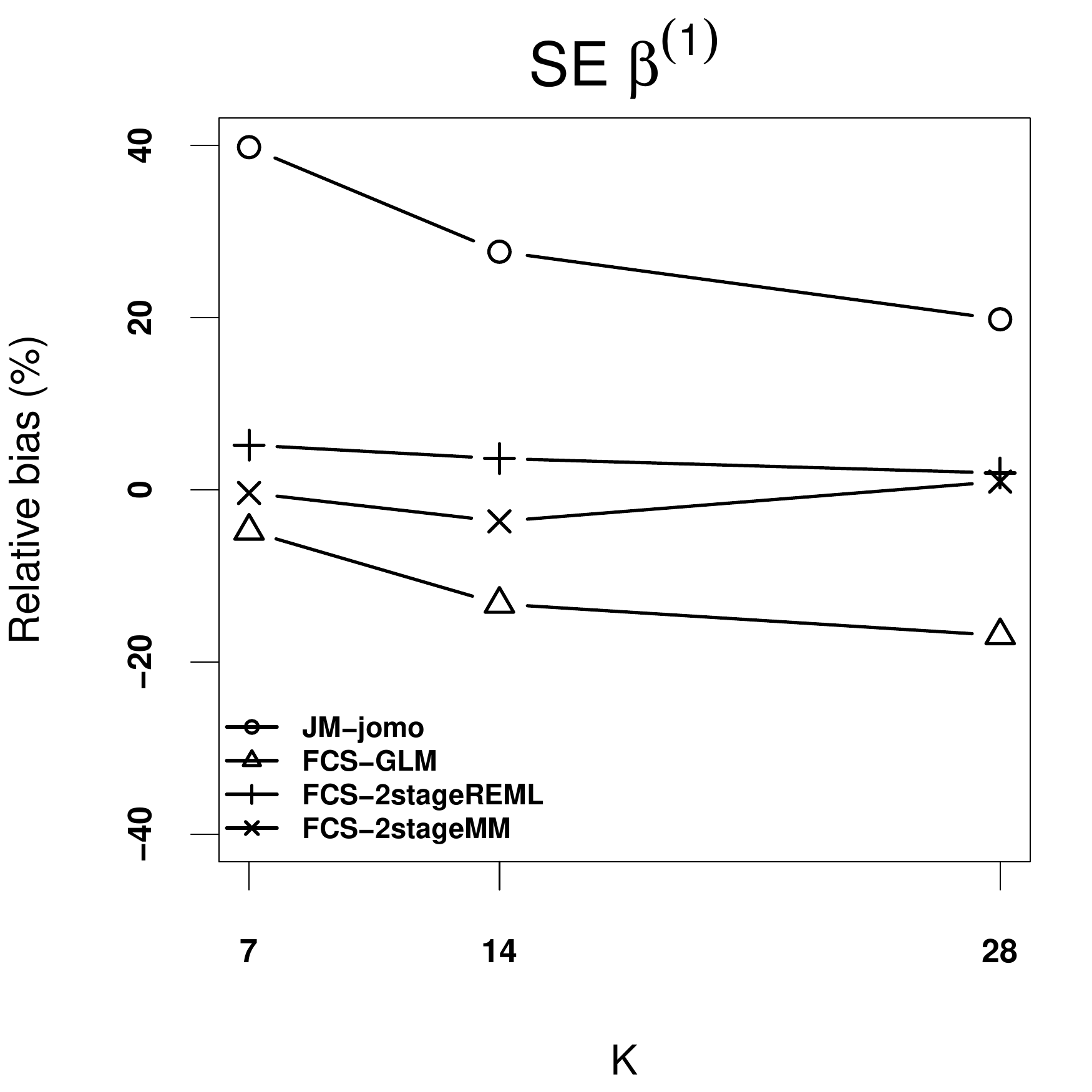} \includegraphics[width=6cm, height=8cm]{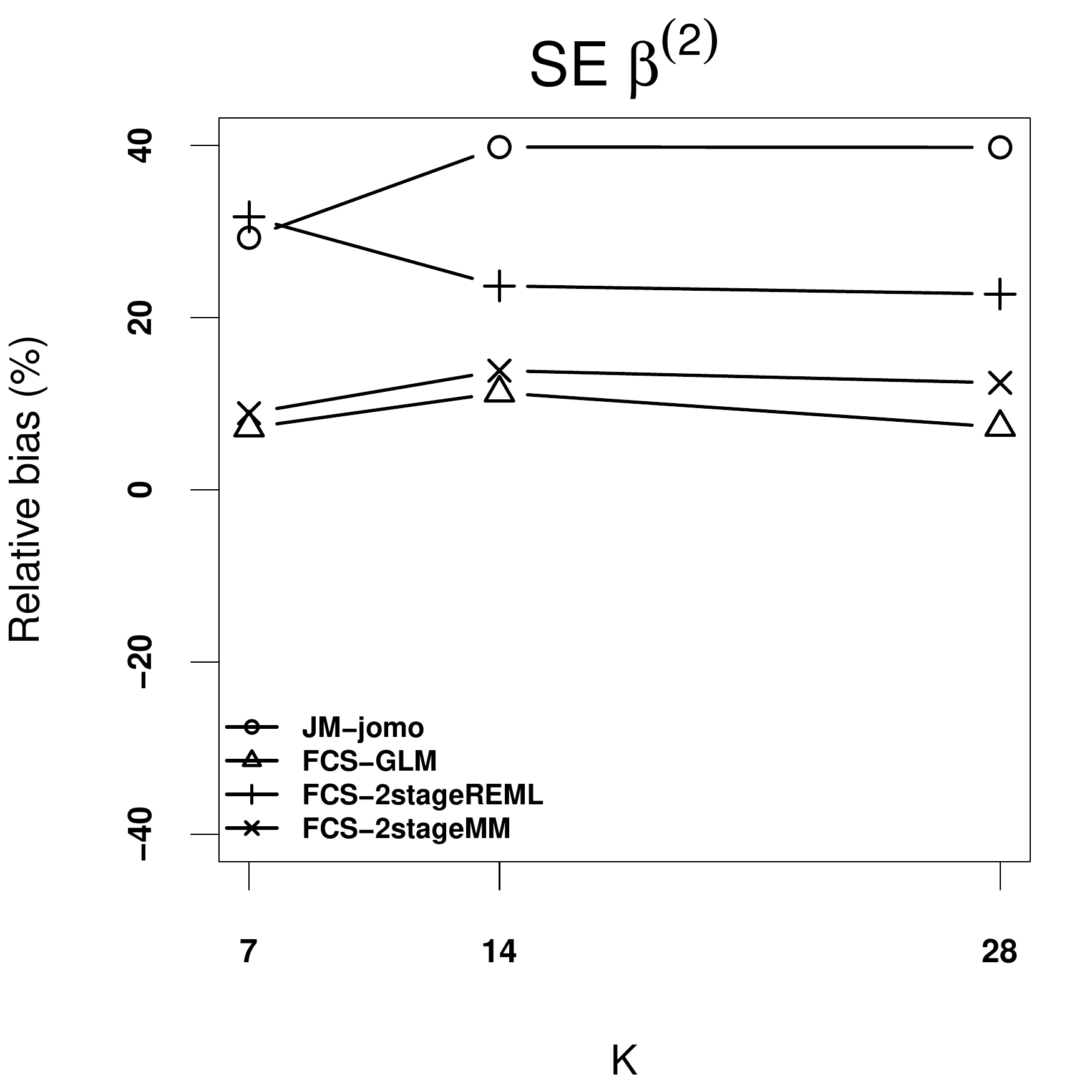}
\end{center}
\end{figure}

\subsubsection{Robustness to the cluster size}\label{Annexnk}
~~
\begin{figure}[H]
\begin{center}
\caption{Robusteness to the cluster size: estimate of the relative bias for the SE estimate for $\widehat{\beta^{(1)}}$ (left), $\widehat{\beta^{(2)}}$ (right) according to $\Nbind_{\nbclust}$ for each MI method.  The estimated relative bias is calculated by the difference between the model SE and the empirical SE, divided by the empirical SE. Criteria are based on 500 incomplete datasets.\label{figa3}}
\includegraphics[width=6cm, height=8cm]{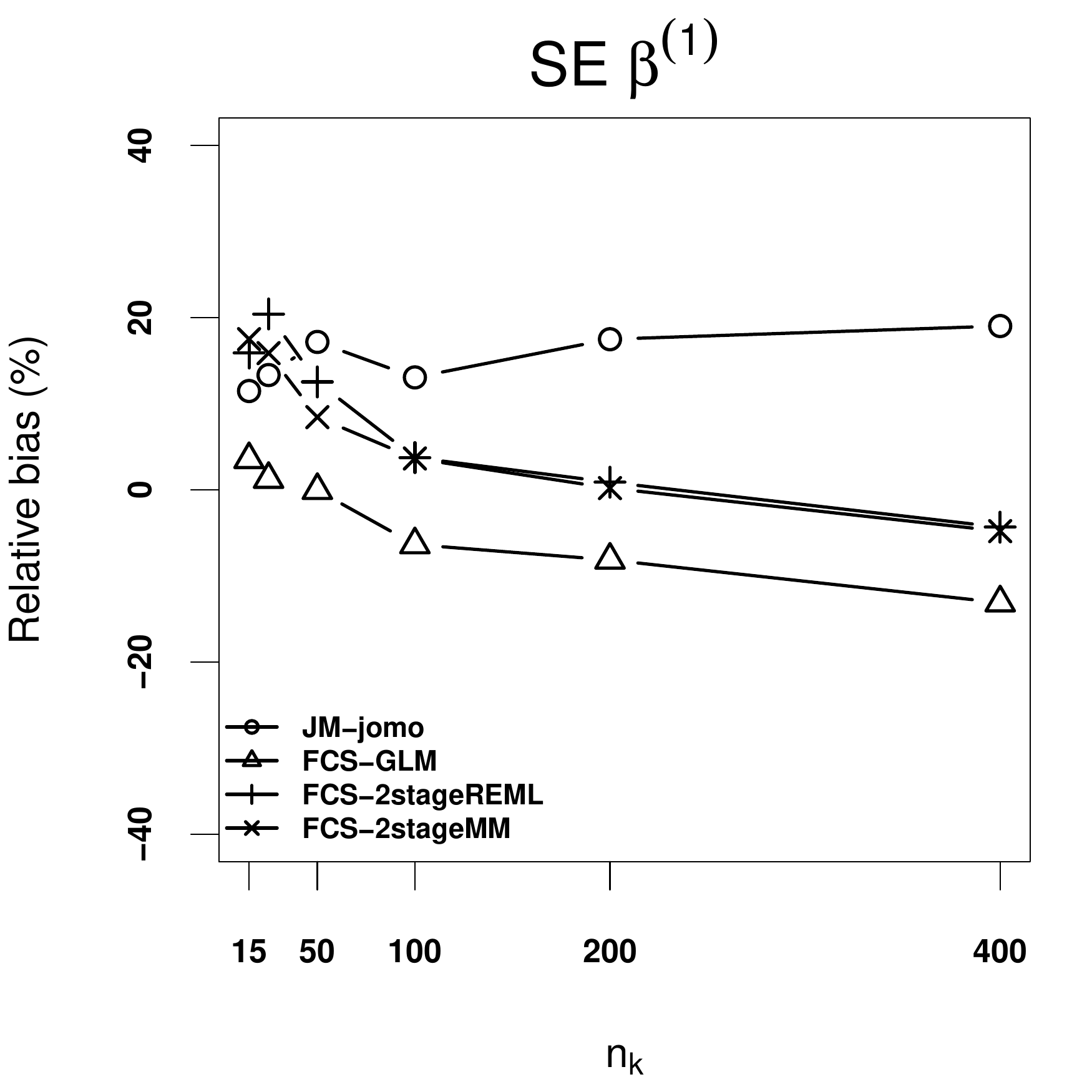} \includegraphics[width=6cm, height=8cm]{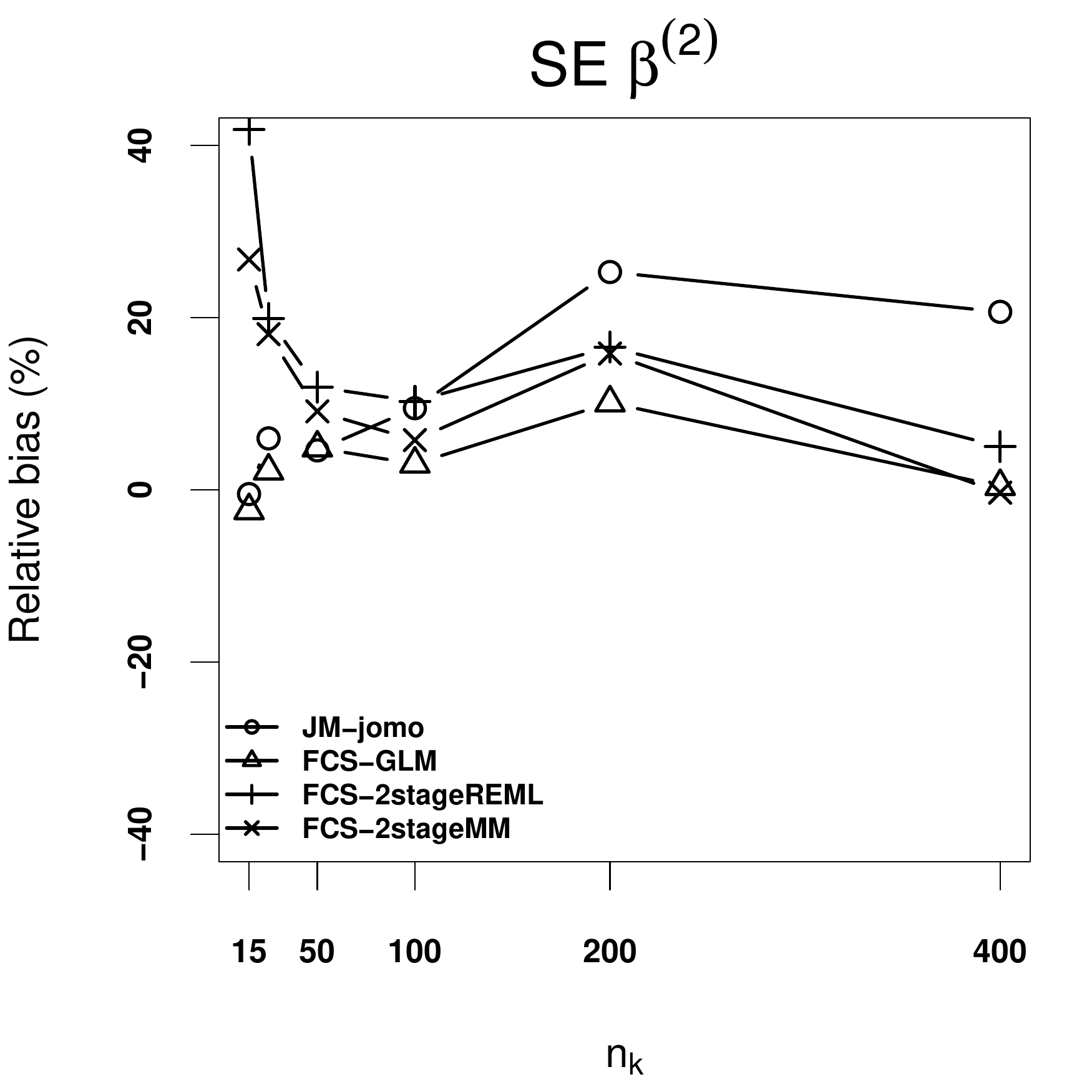}
\end{center}
\end{figure}

\subsubsection{Other configurations}\label{Annexother}
~~
\begin{figure}[H]
\begin{center}
\caption{Distribution of the relative bias over the 19 configurations for several methods (Full, CC, JM-jomo, FCS-GLM, FCS-2stageREML, FCS-2stageMM) and several parameters of interest ($\beta^{(1)}, \beta^{(2)}, \psi_{00}, \psi_{11}$, SE $\beta^{(1)}$, SE $\beta^{(2)}$). One point represents the relative bias observed for one configuration. \label{Figoverall}}
\includegraphics[scale=.85]{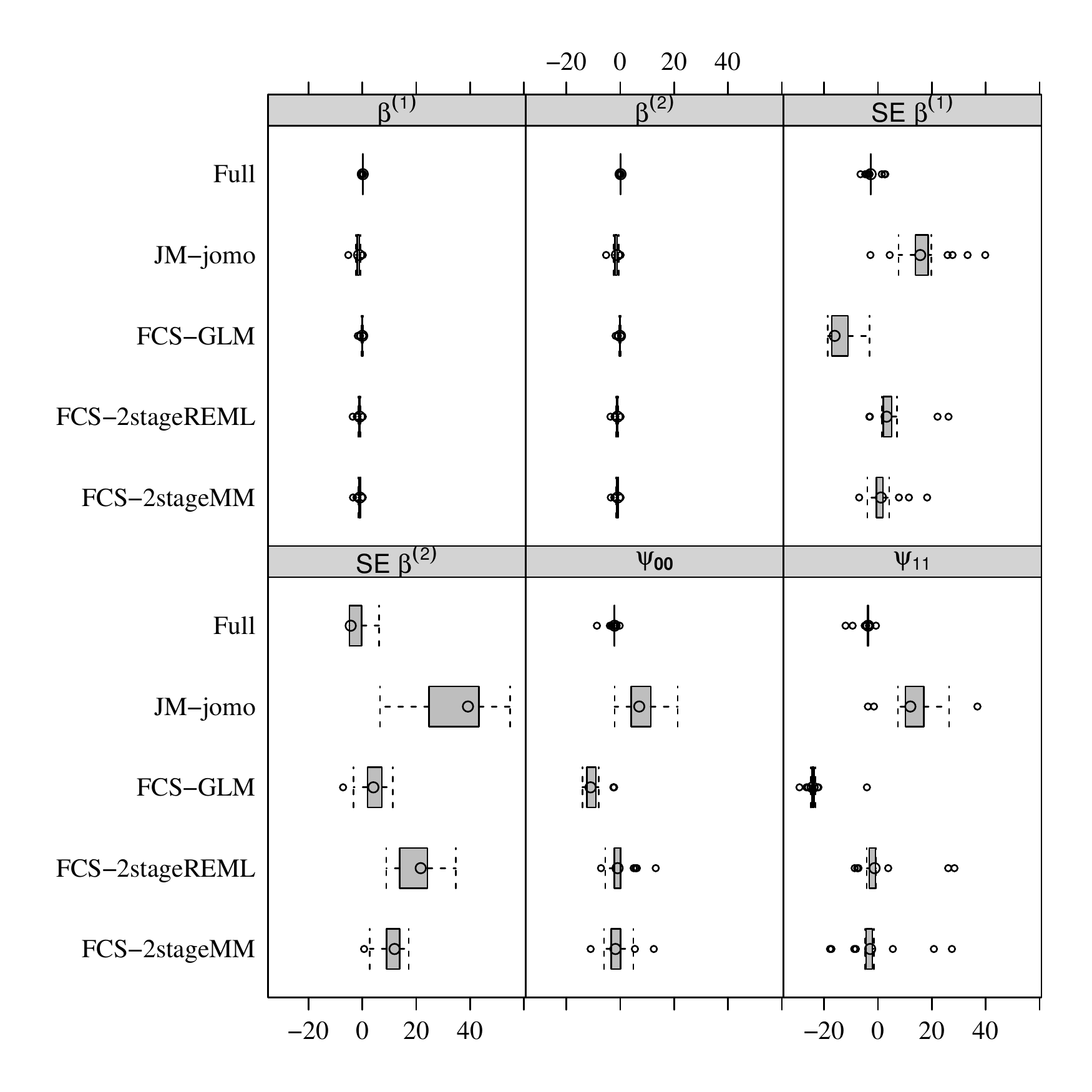}
\end{center}
\end{figure}

\subsubsection{Influence of the random slope}\label{newappendixcong}
\begin{figure}[H]
\begin{center}
\caption{Distribution of the estimate of $\sqrt{\psi_{00}}$ over the 500 generated datasets for Full, CC and JM-jomo methods. Both configurations are considered: a one with a random slope (in grey, corresponding to the base-case configuration) and one without random slope (in blue). The red dashed line represents the true value of  $\sqrt{\psi_{00}}$.\label{newfigcong}}
\includegraphics[scale=.4]{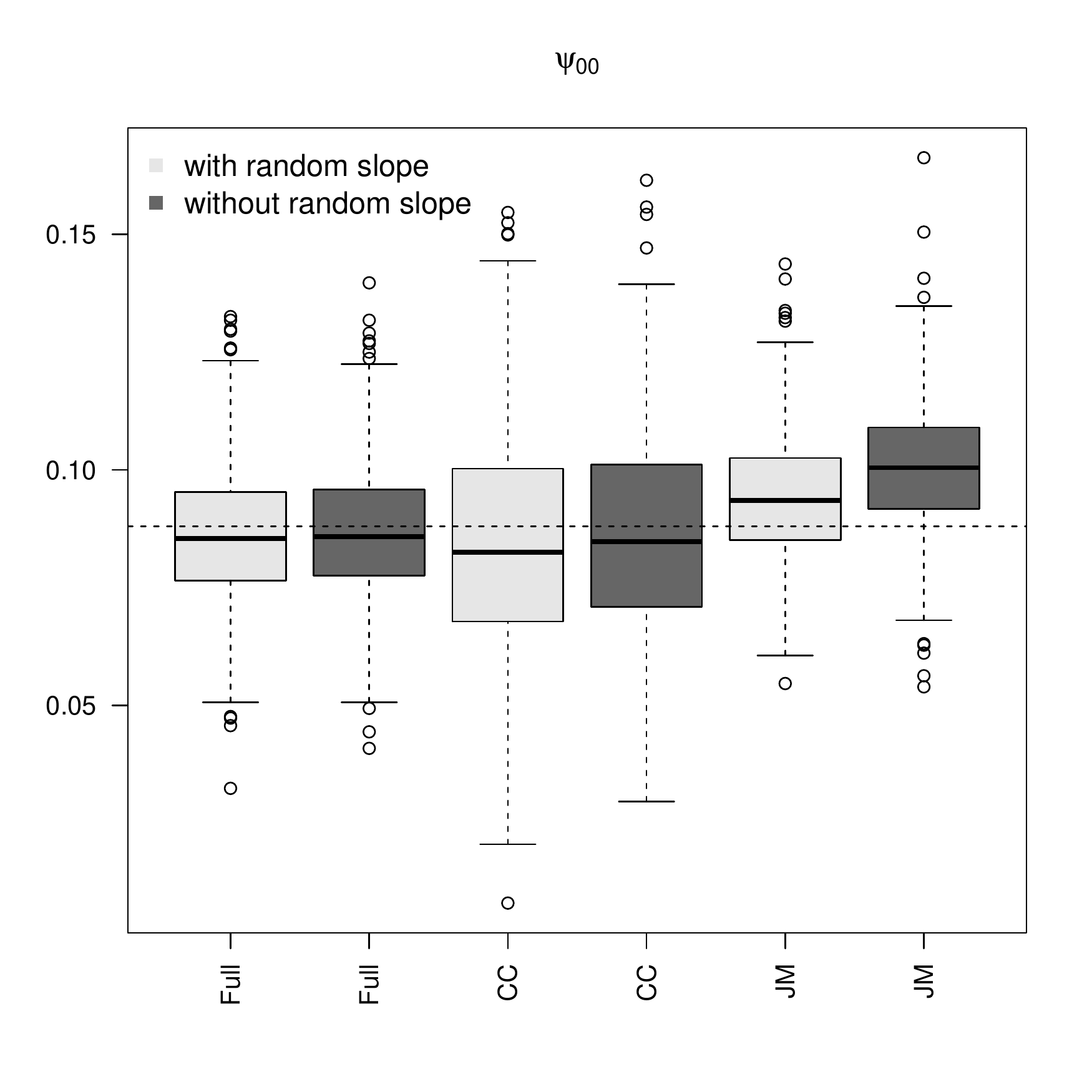}
\end{center}
\end{figure}

The multivariate version of model (\ref{glm}) used in JM-jomo requires that all missing variables are in the left hand side of the model (\ref{glmmulti}). However, if the analysis model includes a random slope in the right hand side (like in the simulation study \ref{simudesign}), then the imputation model is misspecified. To assess the influence of this misspecification on the random effect variance, we compare the estimates of $\psi_{00}$ when the outcome of the model is generated according to an analysis model including a random slope (base-case configuration), with the estimates of $\psi_{00}$ when the outcome of the model is generated according to an analysis model with a random intercept only. Estimates are reported in Figure \ref{newfigcong}.
A bias is observed even if the outcome is generated from a model with no random slope, indicating that misspecification of the imputation model is not the main reason for the observed bias in the base-case configuration. As shown in Section \ref{robvarrand} it is more likely that the use of wrongly informative prior distributions  biases the inference in presence of very small values for the level-2 variances.

\end{document}